%% file: pati-salam-g-2.tex
\documentclass[a4paper,11pt]{article}
\pdfoutput=1 

\usepackage{jheppub} 

\usepackage[T1]{fontenc} 

\usepackage{bm}         
\usepackage{array}
\usepackage{multirow}   
\usepackage{booktabs}   
\usepackage{float}      

\newcommand{\gev}{\,\textrm{GeV}}

\newcommand{\sgn}{\operatorname{sgn}}

\renewcommand{\i}{\mathrm i}
\renewcommand{\(}{\left(}
\renewcommand{\)}{\right)}

\title{A to Z of the Muon Anomalous Magnetic Moment in the MSSM with Pati-Salam at the GUT scale}

\author[a,b]{Alexander S. Belyaev,}
\author[e]{Jos\'e E. Camargo-Molina,}
\author[a]{Steve F. King,}
\author[d]{David J. Miller,}
\author[c,e]{Ant\'onio P. Morais}
\author[a]{and Patrick B. Schaefers}

\affiliation[a]{School of Physics \& Astronomy, University of Southampton, Southampton SO17 1BJ, UK}
\affiliation[b]{Particle Physics Department, Rutherford Appleton Laboratory, Chilton, Didcot, Oxon OX11 0QX, UK}
\affiliation[c]{Departamento de F\'\i sica, Universidade de Aveiro and CIDMA, Campus de Santiago, \linebreak \mbox{3810-183 Aveiro}, Portugal}
\affiliation[d]{SUPA, School of Physics and Astronomy, University of Glasgow,  Glasgow, G12 8QQ, UK}
\affiliation[e]{Department of Astronomy and Theoretical Physics, Lund University, 221 00 Lund, Sweden}

\emailAdd{A.Belyaev@soton.ac.uk}
\emailAdd{Eliel@thep.lu.se}
\emailAdd{S.F.King@soton.ac.uk}
\emailAdd{David.J.Miller@glasgow.ac.uk}
\emailAdd{aapmorais@ua.pt}
\emailAdd{P.Schaefers@soton.ac.uk}

\abstract{We analyse the low energy predictions of the minimal supersymmetric standard model (MSSM) arising from a GUT scale Pati-Salam gauge group further constrained by an 
\mbox{$A_4 \times Z_5$} family symmetry, resulting in four soft scalar masses at the GUT scale:
one left-handed soft mass $m_0$ and three right-handed soft masses $m_1,m_2,m_3$, one for each generation.
We demonstrate that this model, which was initially developed to describe the neutrino sector,  can explain collider and non-collider measurements such as the dark matter relic density, the Higgs boson mass and, in particular, the anomalous magnetic moment of the muon $(g-2)_\mu$. Since about two decades, $(g-2)_\mu$  suffers a puzzling about 3$\,\sigma$ 
excess of the experimentally measured value over the theoretical prediction, 
which our model is able to fully resolve. 
As the consequence of this resolution,
our model predicts specific regions of the parameter space
with the specific properties
including  light smuons and neutralinos, which 
could also potentially explain di-lepton excesses observed by CMS and ATLAS.}


\begin{document} 
\maketitle
\flushbottom

\input{01-intro.tex}            
\input{02-model.tex}            

\input{03-g-2.tex}              
\input{04-constraints.tex}      
\input{05_1-results_inclusive-scan.tex}

\input{05_2-results_small-mu.tex}
\input{05_4-results_large-mu2.tex}
\input{06-vacuum-stability.tex} 
\input{07-conclusions.tex}      


\acknowledgments
The authors acknowledge the use of the IRIDIS High Performance Computing Facility, and
associated support services at the University of Southampton, in the completion of this work.
ASB, SFK and PBS acknowledge partial support from the InvisiblesPlus RISE from the European
Union Horizon 2020 research and innovation programme under the Marie Sklodowska-Curie grant
agreement No 690575. SFK acknowledges partial support from the Elusives ITN from the European
Union Horizon 2020 research and innovation programme under the Marie Sklodowska-Curie grant
agreement No 674896.
AB and SFK acknowledges partial  support from the STFC grant ST/L000296/1.
DJM acknowledges patial  support from the STFC grant ST/L000446/1.
AB also thanks the NExT Institute, Royal Society Leverhulme Trust Senior Research Fellowship LT140094 and
Soton-FAPESP grant for patial  support.
AM is supported by the FCT grant SFRH/BPD/97126/2013 and partially by the H2020-MSCA-RISE-2015 Grant agreement No StronGrHEP-690904,  and  by  the  CIDMA  project UID/MAT/04106/2013. AM also acknowledges the THEP group at Lund University for all hospitality and support provided for the development of this work. JECM wants to thank Ben O'Leary for discussions in the early stages of this work. 


\newpage

\bibliographystyle{JHEP}
\bibliography{bib}

\end{document}

%% file: 01-intro.tex
\section{Introduction}
\label{sec:introduction}

Supersymmetry (SUSY) (for a review see e.g. \cite{Chung:2003fi})
remains an attractive candidate for new physics beyond the Standard Model (SM),
even if there is to date no direct evidence for it  at colliders, most notably the Large Hadron Collider (LHC). However, there remain good motivations for considering SUSY, which are worth repeating, namely that it opens up the possibility for gauge coupling unification, provides a viable dark matter (DM) candidate such as the
R-parity stabilized lightest neutralino, and addresses the big hierarchy problem of the SM. Despite the lack of evidence for SUSY at the LHC, including the lack of non-standard flavour signals in LHCb detector, almost for two decades there remains one stubborn experimental inconsistency in the SM coming from the anomalous magnetic moment of the muon,
which is often overlooked or ignored for one or another reason.
It is well known that SUSY can account for this inconsistency, provided that there are light
sleptons and charginos, which by themselves are not inconsistent with LHC constraints on new coloured particles.
It remains an intriguing question, which we shall address in this paper,
whether this data can be accounted for by a well motivated unified SUSY model 
consistent with other  collider and non-collider constraints including DM.

The magnetic moment of the muon, as predicted by the Dirac equation, is related to the particle's spin $\bm{S}$ by
\begin{align}
\bm{M} = g_{\mu} \dfrac{e}{2 m_{\mu}} \bm{S}
\label{eq:Muon_magnetic}\, ,
\end{align}
where, at classical level, the gyromagnetic ratio is $g_{\mu} = 2$. Small deviations from this value are induced at the quantum level and can be parametrized by the so called anomalous magnetic moment of the muon
\begin{align}
a_{\mu} = \dfrac{g_{\mu} - 2}{2}
\label{eq:amu}\, .
\end{align}
$a_\mu$ is one of the most precisely measured quantities in modern particle physics. The E821 experiment at the Brookhaven National Laboratory has measured $a_\mu$ to $0.54\,$ppm~\cite{Bennett:2006fi,Agashe:2014kda}, resulting in
\begin{align}
	a_\mu^{\rm exp} = 
	11 659 2091(63) \times 10^{-11}.
\end{align}
New experiments at Fermilab~\cite{Grange:2015fou} and J-PARC \cite{Saito:2012zz} promise to improve this accuracy by a factor of four. The SM theory prediction  is of a comparable accuracy (for useful reviews, see \cite{Jegerlehner:2009ry,Blum:2013xva,Benayoun:2014tra,Knecht:2014sea}). This prediction includes QED corrections to five loops~\cite{Aoyama:2012wk} (see also \cite{Kataev:2012kn,Lee:2013sx,Kurz:2013exa,Kurz:2015bia}) as well as weak corrections to two loops~\cite{Czarnecki:2002nt,Gnendiger:2013pva} and hadronic corrections~\cite{Nyffeler:2009tw,Davier:2010nc,Hagiwara:2011af,Benayoun:2012wc,Colangelo:2014dfa,Kurz:2014wya,Colangelo:2014qya,Colangelo:2014pva,Pauk:2014rfa,Colangelo:2015ama,Benayoun:2015gxa} (see also \cite{Blum:2014oka,Jin:2015eua,Chakraborty:2015ugp,Aubin:2015rzx,Blum:2015you} for lattice QCD evaluations). 
The uncertainties in the hadronic corrections, which rely on data for $e^+e^- \to $ hadrons, vary somewhat between authors. In all combinations, there remains a significant tension between experiment and theoretical prediction. This discrepancy ranges from
\begin{align}
\Delta a_\mu = a_\mu^{\rm exp} - a_\mu^{\rm SM}  = 237(86) \times 10^{-11}
\label{eq:Damu_1}
\end{align}
to
\begin{align}
\Delta a_\mu = a_\mu^{\rm exp} - a_\mu^{\rm SM}  = 278(80) \times 10^{-11},
\label{eq:Damu_2}
\end{align}
which are $2.8\,\sigma$ and $3.4\,\sigma$ tensions respectively~\cite{Knecht:2014sea}. In the interest of compatibility with other studies, here we will use the deviation of experiment from the SM prediction quoted in Ref.~\cite{Agashe:2014kda}, which is
\begin{align}
\Delta a_{\mu} = a_\mu^{\rm exp} - a_\mu^{\rm SM}  = 288(80) \times 10^{-11}
\label{eq:Damu_3}\, .
\end{align}

If this discrepancy persists and reaches even higher significance
when confronted with new experiments and/or improvements to the SM hadronic contributions,
it may become a sign of new physics beyond the SM. In particular, within supersymmtric models, the deviation from the SM prediction may be totally or partially attributed to smuon-neutralino and sneutrino-chargino loops~\cite{Grifols:1982vx,Ellis:1982by,Chakrabortty:2013voa,Chakrabortty:2015ika,Barbieri:1982aj,Kosower:1983yw,Yuan:1984ww,Romao:1984pn,Lopez:1993vi,Moroi:1995yh,Martin:2000cr,Czarnecki:2001pv,Cho:2011rk,Endo:2011mc,Endo:2011xq,Endo:2011gy,Evans:2012hg,Endo:2013bba,Mohanty:2013soa,Ibe:2013oha,Akula:2013ioa,Okada:2013ija,Endo:2013lva,Bhattacharyya:2013xma,Gogoladze:2014cha,Kersten:2014xaa,Li:2014dna,Chiu:2014oma,Badziak:2014kea,Calibbi:2015kja,Kowalska:2015zja,Wang:2015rli}. Although $\Delta a_\mu$ may be accommodated in the Minimal Supersymmetric Standard Model (MSSM) (see e.g.\ \cite{Ibe:2013oha,Endo:2013bba}) with its large number of free parameters, finding a suitable value in more constrained supersymmetric models can be challenging. For example, in the well studied Constrained MSSM (CMSSM), in which the supersymmetric soft-breaking masses are given common values at some high energy scale, it is difficult to achieve the desired value of $a_\mu$~\cite{Bechtle:2012zk,Buchmueller:2012hv,Balazs:2013qva}. Of course, if one is willing to attribute only part of the discrepancy to supersymmetric effects, then simple models of Grand Unification that satisfy all constraints become viable (see e.g.\ \cite{Miller:2013jra,Miller:2014jza}) but are no more attractive for explaining $a_\mu$ than the SM. 

Another possible class of models which could address $(g-2)_{\mu}$
are the SUSY GUT models with normal mass hierarchy with non-universal
scalar masses for the first two and the third generation of sfermions~\cite{Baer:2004xx}.
Also, the $(g-2)_{\mu}$ problem can be addressed in the essentialy non-universal 
model such as  the phenomenological MSSM (pMSSM) scenario
\cite{Djouadi:1998di} which is based on the following simplifying assumptions: 
\begin{itemize}
\item First and second generation universality for low energy soft masses 
$m_{Q_1},m_{U_1},m_{D_1},m_{L_1},m_{E_1}$ (equal to $m_{Q_2},m_{U_2},m_{D_2},m_{L_2},m_{E_2}$,
respectively)
\item Separate low energy soft masses for third generation scalar masses 
$m_{Q_3},m_{U_3},m_{D_3},m_{L_3},m_{E_3}$
\item Separate low energy gaugino masses $M_1,M_2,M_3$
\item Separate trilinear parameters $A_t,A_b,A_{\tau}$
\end{itemize}

In this paper we will investigate contributions to $a_\mu$ that arise from a
conceptually  different MSSM model 
based on a high energy (GUT scale) Pati-Salam gauge group combined with an $A_4 \times Z_5$ family symmetry~\cite{King:2014iia}.
The point is that this model was initially motivated not by $(g-2)_\mu$
but by the fact that it provides an excellent description of quark and lepton masses, mixing and CP violation. The model predicts the following high energy (GUT scale) soft mass parameters:
\begin{itemize}
\item A universal high energy soft scalar mass for all left-handed squarks and sleptons of all three families, $m_0$
(i.e. $m_{Q_i}$ and $m_{L_i}$ are unified into $m_0$ at the GUT scale )
\item Three high energy soft mass parameters for the right-handed squarks and leptons, one for each 
family $m_1$, $m_2$, $m_3$ (i.e. $m_{U_i}$, $m_{D_i}$ and $m_{E_i}$ are unified into $m_i$ at the GUT scale,
respectively for $i=1,2,3$ )
\item Separate high energy gaugino masses $M_1,M_2,M_3$
\item Separate trilinear parameters $A_t,A_b,A_{\tau}$
\end{itemize}
These soft mass boundary conditions are consistent with the (s)particle groupings 
dictated by the model as shown in figure \ref{A2Z}.
We will show that this model has also a great potential
to predict $a_\mu$ that is in agreement with the experimental value, while simultaneously providing a viable Dark Matter candidate, maintaining vacuum stability and remaining consistent with all experimental constraints.

In section~\ref{sec:the-model} we will describe the model in some detail, and in section~\ref{sec:g-2} we clarify the leading contributions to $\Delta a_\mu$. We will discuss constraints from experiment, including collider constraints and those on the Dark Matter relic density, in section~\ref{sec:constraints}. We present our results, including some example scenarios, in section~\ref{sec:results}. Finally we investigate vacuum stability for these example scenarios in section~\ref{sec:vacuum-stability}, before concluding in section~\ref{sec:conclusion}.

%% file: 02-model.tex
\section{The Model}
\label{sec:the-model}
\begin{figure}[tbp]
\centering
\includegraphics[width=0.4\textwidth]{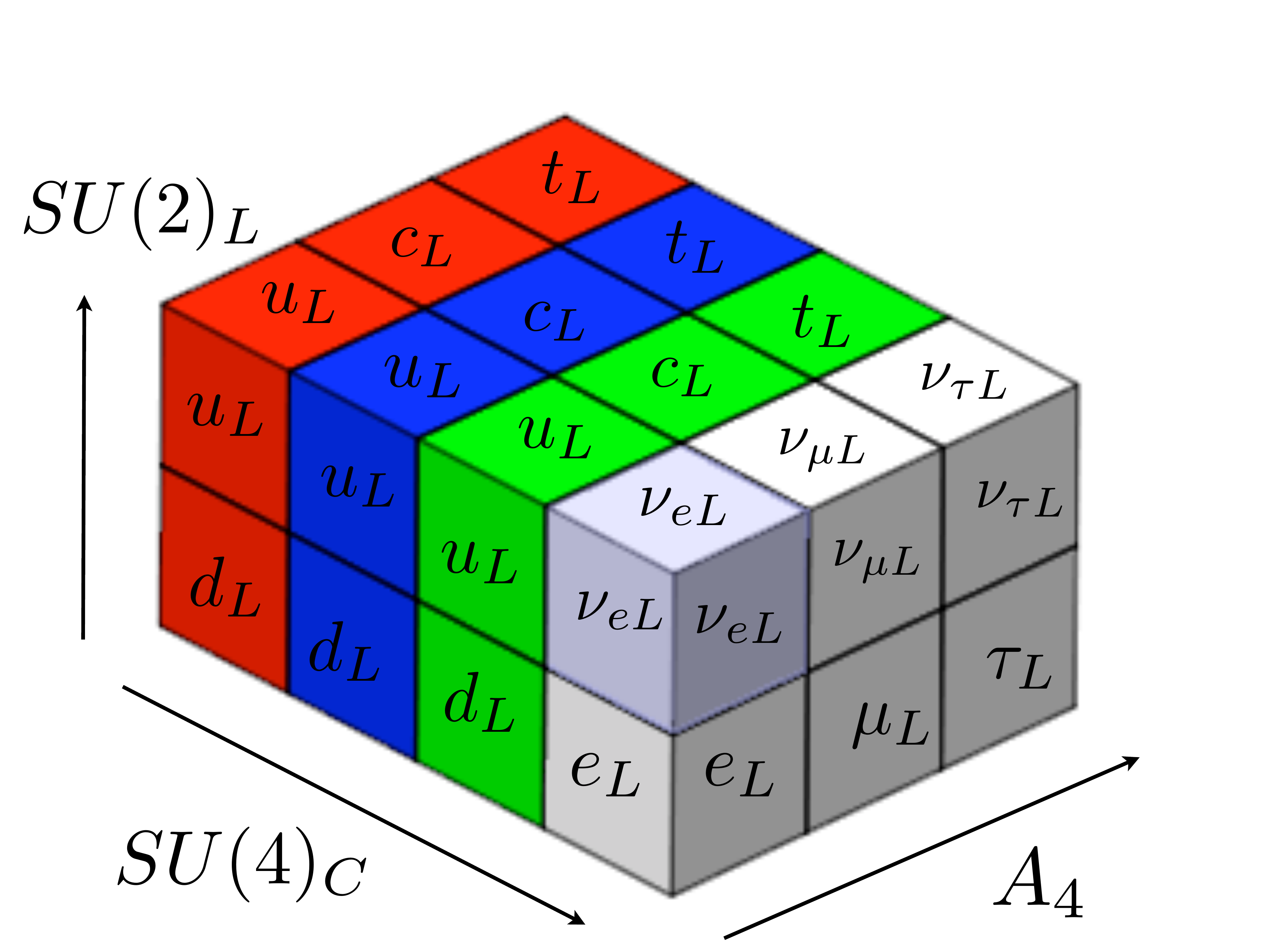}
\includegraphics[width=0.4\textwidth]{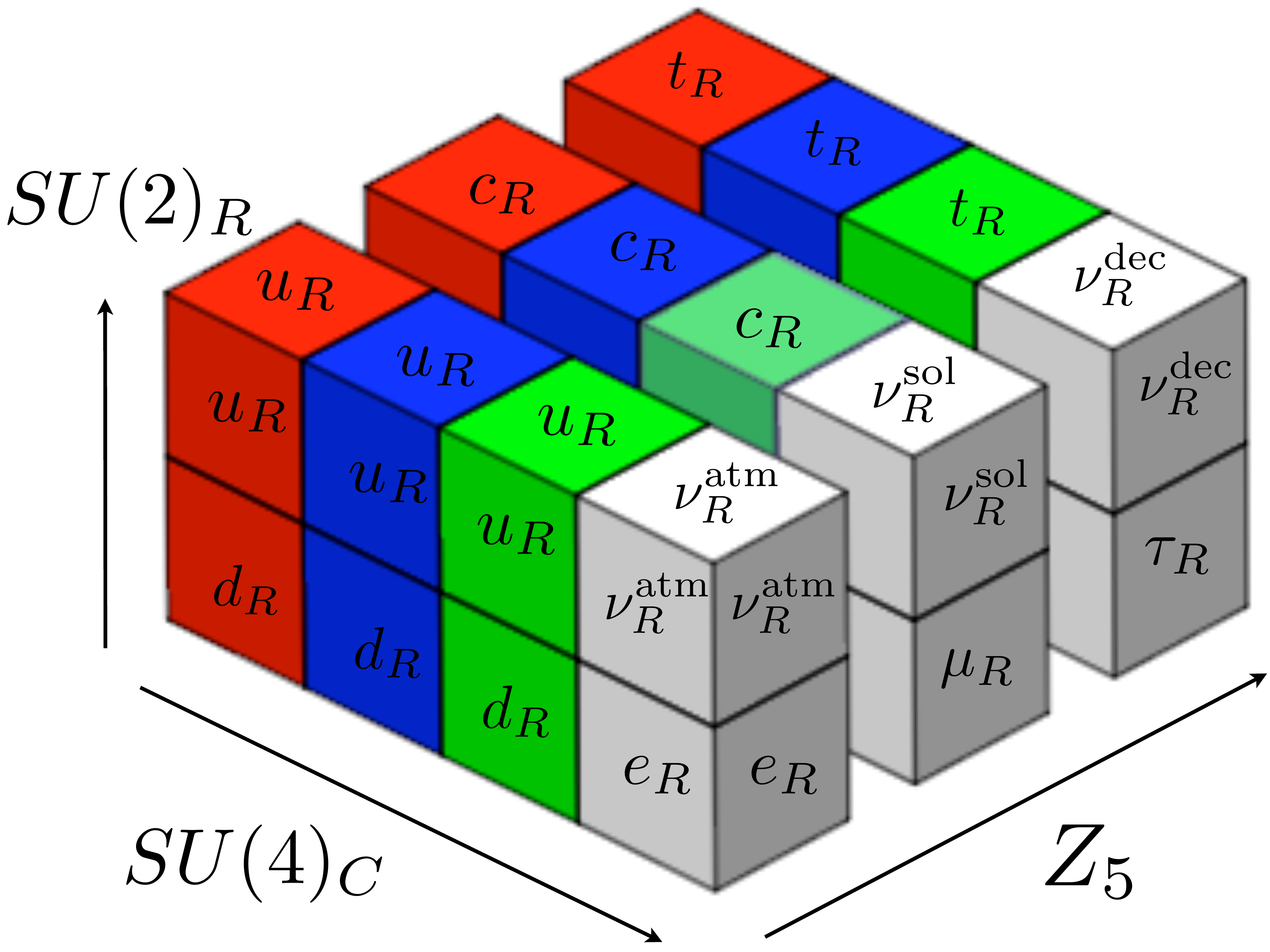}
\vspace*{-4mm}
    \caption{$A$ to $Z$ of flavour with Pati-Salam, where $A\equiv A_4$ and $Z\equiv Z_5$. 
         The left-handed families form a triplet of $A_4$ and are
         doublets of $SU(2)_L$.
        The right-handed families are distinguished by $Z_5$
and are doublets of $SU(2)_R$.
The $SU(4)_C
$ unifies the quarks and leptons with leptons as the fourth colour,
   depicted here as white.
     } \label{A2Z}
\vspace*{-2mm}
\end{figure}

An ``A to Z of flavour with Pati-Salam'' based on the Pati-Salam gauge group has been proposed 
\cite{King:2014iia} as sketched in figure \ref{A2Z}.
The Pati-Salam symmetry leads to $Y^u=Y^{\nu}$, where the columns of the Yukawa matrices 
are determined by flavon alignments.
The first column is proportional to the alignment $(0,e,e)$,
the second column proportional to the orthogonal alignment
$(a,4a,2a)$, and the third column is proportional to the alignment $(0,0,c)$,
where $e\ll a\ll c$ gives the hierarchy $m_u\ll m_c\ll m_t$.
This structure predicts a Cabibbo angle 
$\theta_C\approx 1/4$
in the diagonal $Y^d\sim Y^e$ basis enforced by the 
first three alignments.
It also predicts a normal neutrino mass hierarchy with 
$\theta_{13}\approx 9^{\circ}$, $\theta_{23}\approx 45^{\circ}$
and $\delta \approx 260^{\circ}$ \cite{King:2014iia}.


The model is based on the Pati-Salam (PS) gauge group, 
with $A_4\times Z_5$ (A to Z) family symmetry,
\begin{equation}
SU(4)_{C} \times SU(2)_L \times SU(2)_R\times A_4 \times Z_5.
\label{422A4Z5}
\end{equation}
The quarks and leptons are unified in the PS representations as follows,
\begin{equation}
\begin{aligned}
F &= (4,2,1)_i = \left(\begin{array}{cccc}u&u&u&\nu\\
d&d&d&e\end{array}\right)_i \rightarrow  (Q_i,L_i), \\ 
F^c_i &= (\bar{4},1,2)_i = 
\left(\begin{array}{cccc}u^c&u^c&u^c&\nu^c\\
d^c&d^c&d^c&e^c\end{array}\right)_i \rightarrow (u^c_i,d^c_i,\nu^c_i,e^c_i ),
\label{ql}
\end{aligned}
\end{equation}
\noindent where the SM multiplets $Q_i,L_i,u^c_i,d^c_i,\nu^c_i,e^c_i$ 
resulting from PS breaking are also shown and the 
subscript $i\ (=1,2,3)$ denotes the family index.
The left-handed quarks and leptons form an $A_4$ triplet $F$, 
while the three (CP conjugated) right-handed fields $F^c_i$ are $A_4$ singlets, distinguished by $Z_5$ charges $\alpha, \alpha^3,1$, for $i=1,2,3$, respectively.
Clearly the Pati-Salam model cannot be embedded into an $SO(10)$ Grand Unified Theory (GUT)
since different components 
of the 16-dimensional representation of $SO(10)$ would have to transform differently under $A_4\times Z_5$,
which is impossible, but the PS gauge group and $A_4$ could emerge directly from
string theory.

In the SUSY theory at the GUT scale, from \eqref{ql} there are therefore four different matter
multiplets: $F, F^c_1,F^c_2,F^c_3$, corresponding to the left-handed block and the three
distinct
right-handed blocks
in figure \ref{A2Z} respectively. The GUT-scale scalar soft mass of $F$ will be called $m_0$, while the soft masses of 
$F^c_1,F^c_2,F^c_3$ will be denoted $m_1,m_2,m_3$, respectively,
as discussed in the introduction. The model therefore provides novel
SUSY boundary conditions  for soft masses at the GUT scale, more constrained than the general
MSSM, but less so than the CMSSM. As we shall see, this allows us to account for the 
experimentally observed g-2 of the muon, and will lead to a distinctive and novel low energy
superpartner mass spectrum, with characteristic signatures at the LHC.

The Pati-Salam gauge group is broken at the GUT scale to the SM gauge group,
\begin{equation}
SU(4)_{C}\times SU(2)_L \times SU(2)_R\rightarrow SU(3)_C\times SU(2)_L \times U(1)_Y,
\end{equation}
by PS Higgs, 
$H^c$  and $\overline{H^c}$,
\begin{equation}
\begin{aligned}
{H^c} & = & (u^c_H,d^c_H,\nu^c_H,e^c_H ) \in (\bar{4},1,2), \\ 
 {\overline{H}^c} & = &  (\bar{u}^c_H,\bar{d}^c_H,\bar{\nu}^c_H,\bar{e}^c_H ) \in (4,1,2).
\end{aligned}
\end{equation}
These acquire vacuum expectation values (VEVs) in the ``right-handed neutrino'' directions, with equal VEVs
close to the GUT scale $2\times 10^{16}$ GeV,
\begin{equation}
 \langle {H^c}\rangle = \langle {\nu^c_H}\rangle
 =\langle{\overline{H^c}}\rangle=\langle{\bar{\nu}^c_H}\rangle \sim 2\times 10^{16} \ {\rm GeV},
\label{PS}
\end{equation}
so as to maintain supersymmetric gauge coupling unification.

The model will involve Higgs bi-doublets of two kinds, $h_u$ which lead to up-type quark and neutrino Yukawa couplings and $h_d$ which lead to down-type quark and charged lepton Yukawa couplings.
In addition a Higgs bidoublet $h_3$, which is also an $A_4$ triplet, is used to give the third family
Yukawa couplings. 
After the PS and $A_4$ breaking, most of these Higgs bi-doublets 
will get high scale masses and will not appear in the low energy spectrum. In fact only two
light Higgs doublets will survive down to the TeV scale, namely $H_u$ and $H_d$.
The light Higgs doublet $H_u$ with hypercharge $Y=+1/2$, which couples to up-type quarks and neutrinos,
is a linear combination of components of the Higgs bi-doublets of the kind $h_u$ and $h_3$,
while the light Higgs doublet $H_d$ with hypercharge $Y=-1/2$, 
which couples to down-type quarks and charged leptons,
is a linear combination of components of Higgs bi-doublets of the kind $h_d$ and $h_3$,
\begin{equation}
h_u,h_3 \rightarrow H_u, \ \ \ \ 
h_d,h_3 \rightarrow H_d.
\label{H}
\end{equation}

Therefore, below the GUT scale, the model reduces to the usual MSSM, but with GUT scale boundary conditions
for soft scalar masses as discussed above.

%% file: 03-g-2.tex
\section{\texorpdfstring{One-loop contributions to $\Delta a_{\mu}$}{One-loop contributions to Delta a\_mu}}
\label{sec:g-2}

\begin{figure}[t!]
\centering
\includegraphics[width=0.45\textwidth]{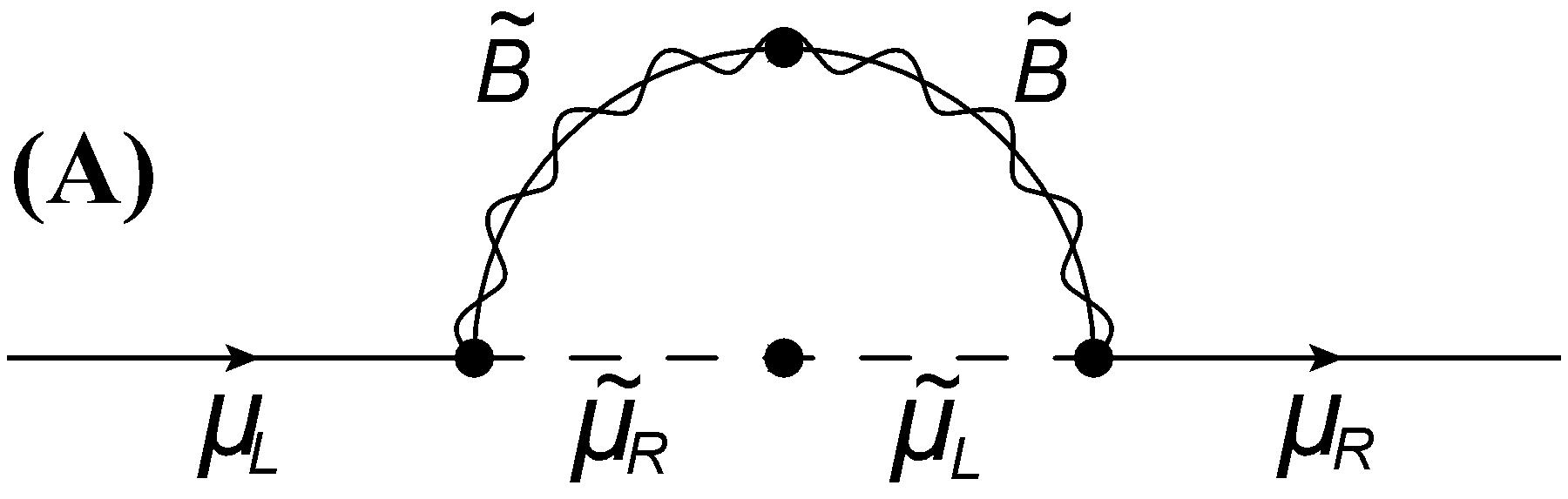}
\includegraphics[width=0.45\textwidth]{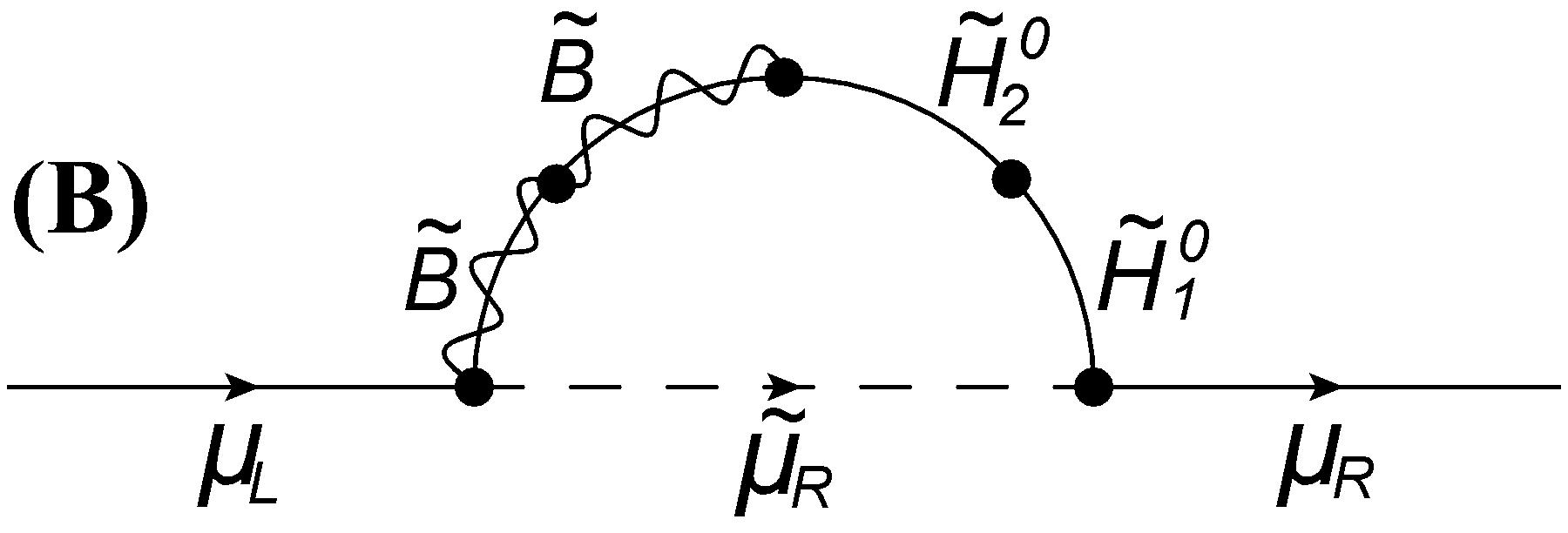}
\includegraphics[width=0.45\textwidth]{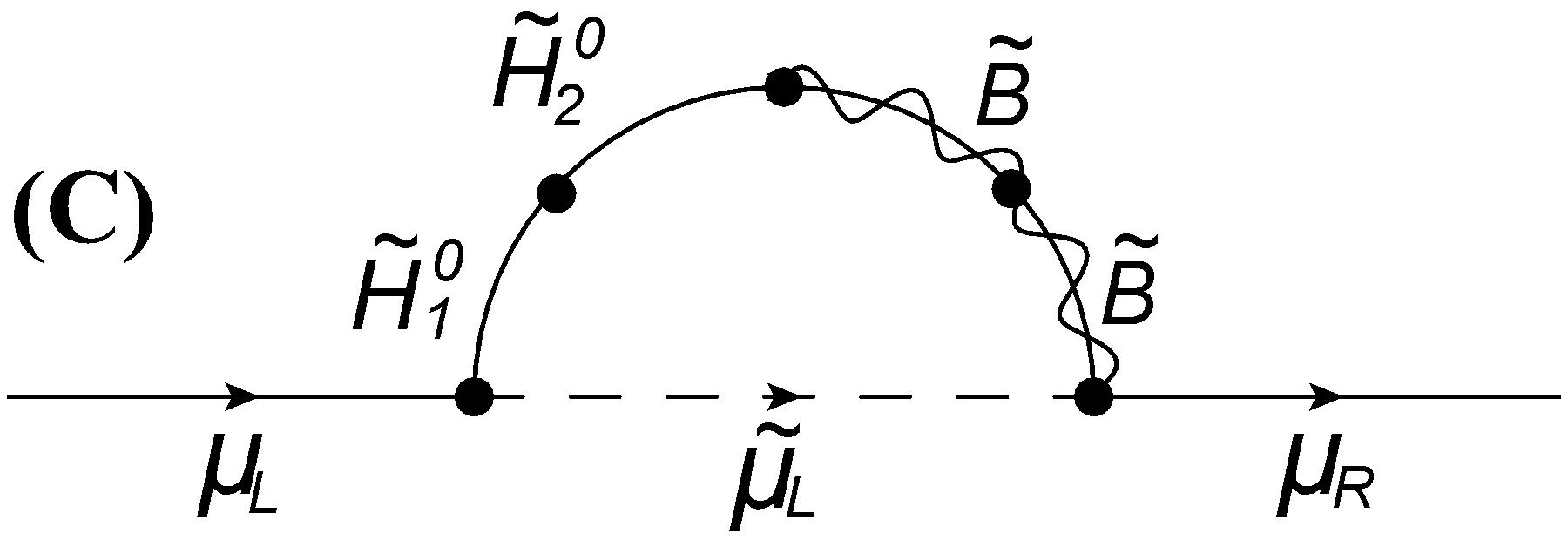}
\includegraphics[width=0.45\textwidth]{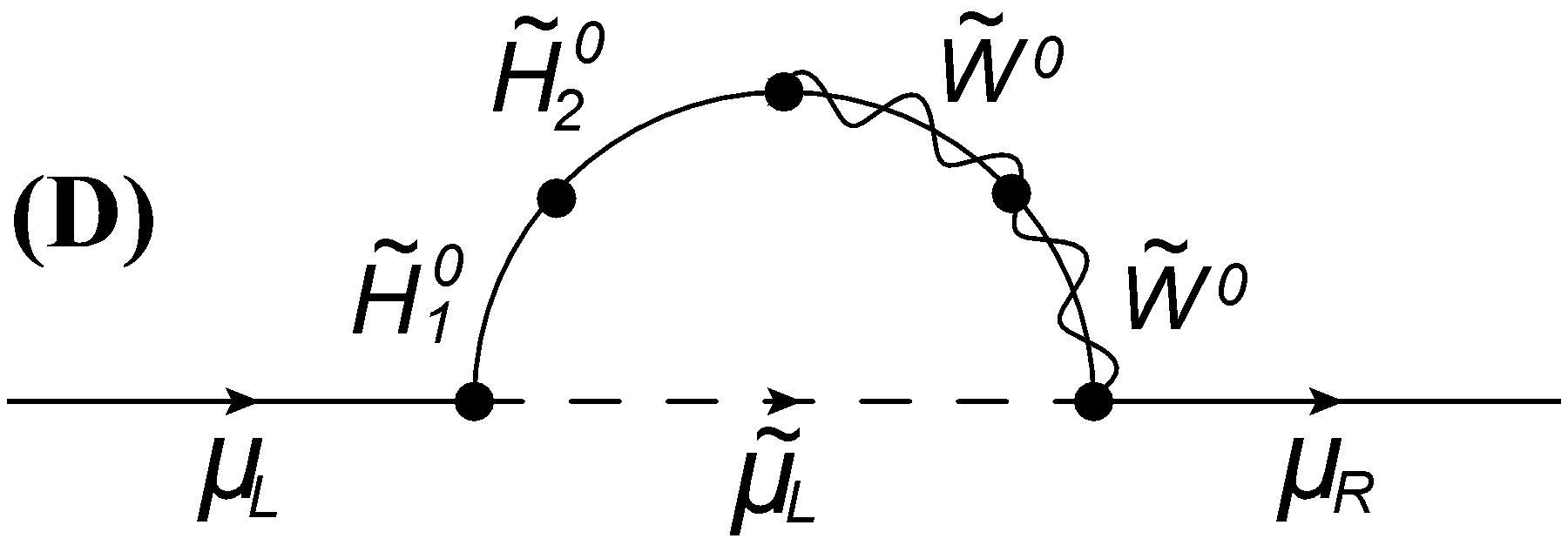}
\includegraphics[width=0.45\textwidth]{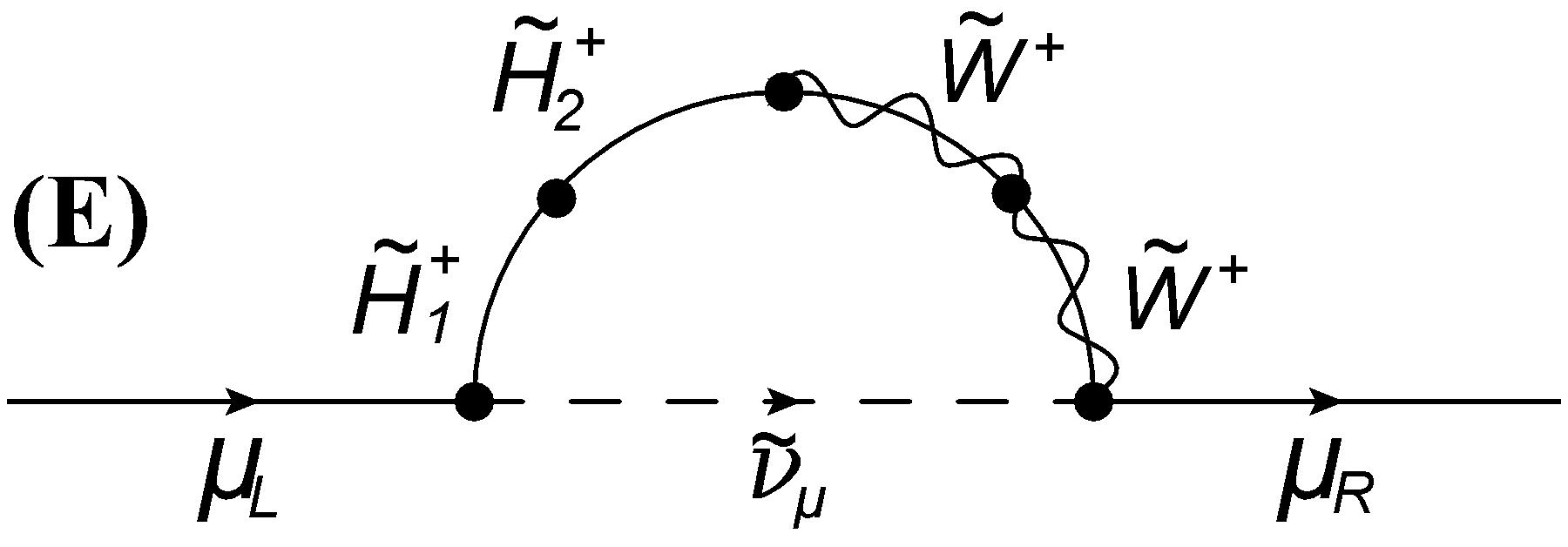}
\caption{One-loop contributions to the anomalous magnetic moment of the muon for supersymmetric models with low-scale MSSM.}
\label{fig:1-loop-amu}
\end{figure}

\begin{figure}[htb!]
\centering
\includegraphics[width=0.495\textwidth]{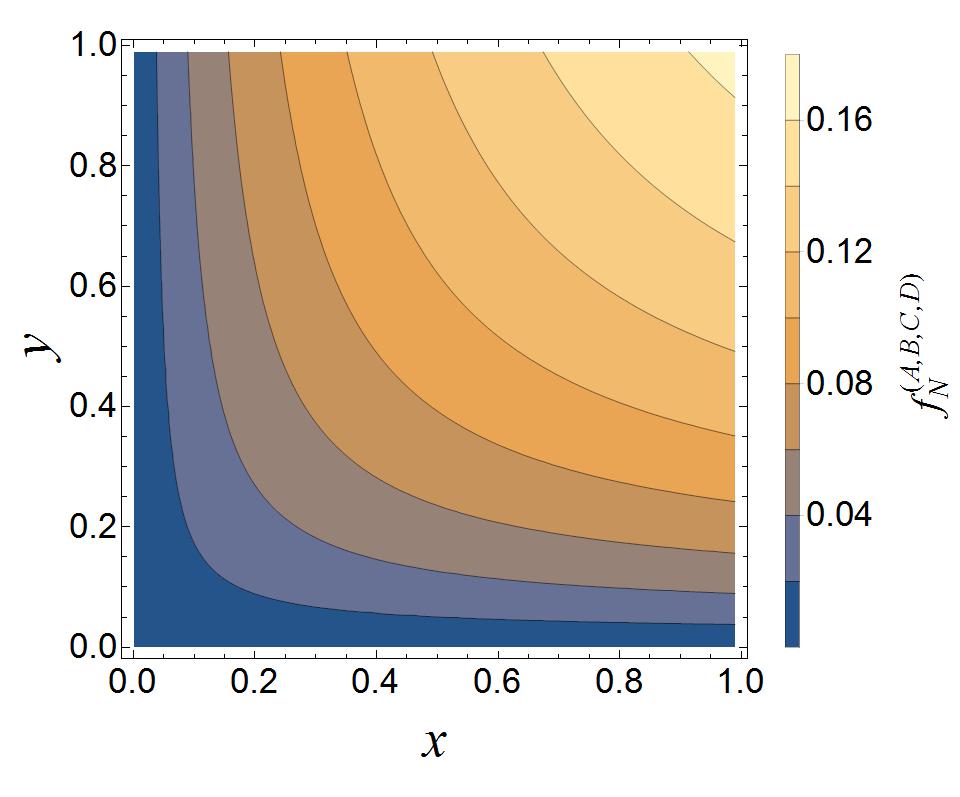}
\includegraphics[width=0.495\textwidth]{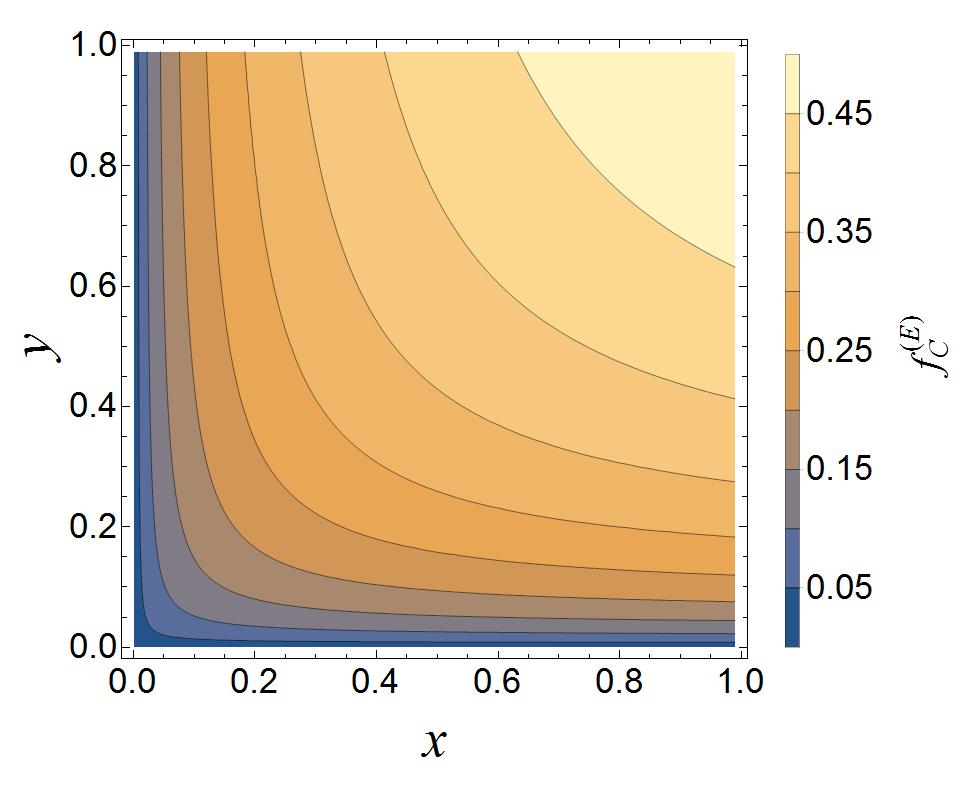}
\caption{Contour plots for $f_{\rm N}\left(x,y \right)$ (left) and $f_{\rm C}\left(x,y \right)$ (right).}
\label{fig:contour}
\end{figure}

The magnetic moment of a massive charged particle is a result of the interaction of its spin with the electromagnetic field. At zeroth order in perturbation theory, the gyromagnetic ratio is predicted to be 2 for every massive particle with semi-integer spin. Deviations from this classical value emerge at the loop-level, where besides SM corrections, new physics contributions may also be relevant. This is indeed the case for the anomalous magnetic moment of the muon, where one-loop supersymmetric contributions are represented in the Feynman diagrams of figure \ref{fig:1-loop-amu}.
These diagrams were computed in \cite{Moroi:1995yh,Endo:2013bba} and give contributions 
\begin{subequations}
\label{eq:loops}
\begin{align}
\Delta a_{\mu}^{(A)}&=\left(\dfrac{M_{1}\mu}{m_{\tilde{\mu}_{L}}^{2}m_{\tilde{\mu}_{R}}^{2}}\right)\dfrac{\alpha_{1}}{4\pi}m_{\mu}^{2}\tan\beta\cdot f^{(A)}_{{\rm N}}\left(\dfrac{m_{\tilde{\mu}_{L}}^{2}}{M_{1}^{2}},\dfrac{m_{\tilde{\mu}_{R}}^{2}}{M_{1}^{2}}\right)\, , \label{eq:loops:A} \\  
\Delta a_{\mu}^{(B)}&=-\left(\dfrac{1}{M_{1}\mu}\right)\dfrac{\alpha_{1}}{4\pi}m_{\mu}^{2}\tan\beta\cdot f^{(B)}_{{\rm N}}\left(\dfrac{M_{1}^{2}}{m_{\tilde{\mu}_{R}}^{2}},\dfrac{\mu^{2}}{m_{\tilde{\mu}_{R}}^{2}}\right)\, , \label{eq:loops:B} \\
\Delta a_{\mu}^{(C)}&=\left(\dfrac{1}{M_{1}\mu}\right)\dfrac{\alpha_{1}}{8\pi}m_{\mu}^{2}\tan\beta\cdot f^{(C)}_{{\rm N}}\left(\dfrac{M_{1}^{2}}{m_{\tilde{\mu}_{L}}^{2}},\dfrac{\mu^{2}}{m_{\tilde{\mu}_{L}}^{2}}\right)\, , \label{eq:loops:C} \\
\Delta a_{\mu}^{(D)}&=-\left(\dfrac{1}{M_{2}\mu}\right)\dfrac{\alpha_{2}}{8\pi}m_{\mu}^{2}\tan\beta\cdot f^{(D)}_{{\rm N}}\left(\dfrac{M_{2}^{2}}{m_{\tilde{\mu}_{L}}^{2}},\dfrac{\mu^{2}}{m_{\tilde{\mu}_{L}}^{2}}\right)\, , \label{eq:loops:D} \\
\Delta a_{\mu}^{(E)}&=\left(\dfrac{1}{M_{2}\mu}\right)\dfrac{\alpha_{2}}{4\pi}m_{\mu}^{2}\tan\beta\cdot f^{(E)}_{{\rm C}}\left(\dfrac{M_{2}^{2}}{m_{\tilde{\nu}_{\mu}}^{2}},\dfrac{\mu^{2}}{m_{\tilde{\nu}_{\mu}}^{2}}\right)\, , \label{eq:loops:E}
\end{align}
\end{subequations}
with $\alpha_1$ and $\alpha_2$ the $U(1)_{Y}$ and $SU(2)_L$ fine structure constants respectively. The functions $f^{(A,B,C,D)}_{{\rm N}}\left(x,y\right)$ and $f^{(E)}_{{\rm C}}\left(x,y\right)$ are given by
\begin{subequations}
\label{eq:f}
\begin{align}
f^{(A,B,C,D)}_{{\rm N}}\left(x,y\right)&=xy\left[\dfrac{-3+x+y+xy}{\left(x-1\right)^{2}\left(y-1\right)^{2}}+\dfrac{2x\log x}{\left(x-y\right)\left(x-1\right)^{3}}-\dfrac{2y\log y}{\left(x-y\right)\left(y-1\right)^{3}}\right]\, , \label{eq:f:1} \\
f^{(E)}_{{\rm C}}\left(x,y\right)&=xy\left[\dfrac{5-3\left(x+y\right)+xy}{\left(x-1\right)^{2}\left(y-1\right)^{2}}-\dfrac{2\log x}{\left(x-y\right)\left(x-1\right)^{3}}+\dfrac{2\log y}{\left(x-y\right)\left(y-1\right)^{3}}\right]\, , \label{eq:f:2}
\end{align}
\end{subequations}
where we use the superscripts $(A,B,C,D)$ and $(E)$ as a short notation to allow omission of the mass ratio arguments.
As described in \cite{Endo:2013bba}, the loop-functions $f^{(A,B,C,D)}_{{\rm N}}$ and $f^{(E)}_{{\rm C}}$ are monotonically increasing for both $x$ and $y$ and are defined in $0 \leq f_{{\rm N, C}} \leq 1 $. From \eqref{eq:loops}, we see that the size of each $\Delta a_{\mu}^{(i)}$ contribution is largely governed by the pre-factor between brackets on the RHS. Therefore, a large $\mu$ combined with light smuons enhances $\Delta a_{\mu}$ via diagram $(A)$ in figure \ref{fig:1-loop-amu}, while keeping the remaining contributions suppressed. However this solution is not unique and in the limit of small $\mu$ the size of the functions $f^{(A,B,C,D)}_{\rm N}$ and $f^{(E)}_{\rm C}$ themselves may distinguish the dominant contributions among diagrams $(B)$ to $(E)$. In particular, we see from the contour plots of figure \ref{fig:contour}, that for a fixed $(x,y)$, say $x \sim y \sim 0.2$, $f^{(E)}_{\rm C} \sim 0.2$ is approximately one order of magnitude larger than $f^{(A,B,C,D)}_{\rm N} \sim 0.02$. We will see in section~\ref{sec:results} the importance of these functions for the explanation of $\Delta a_\mu$.

%% file: 04-constraints.tex
\section{Experimental Constraints}
\label{sec:constraints}
Any successful high-energy completion of the SM should satisfy all known low-energy experimental constraints. In
particular, we require our scenarios to conform to measurements of the Dark Matter (DM) relic density and obey
constraints from the direct detection of DM. The current combined best fit of the DM relic density to data from
\textsc{Planck} and \textsc{Wmap} is $\Omega h^2 = 0.1198 \pm 0.0026$ \cite{Ade:2013zuv}. We will also consider smaller
values of $\Omega h^2$, allowing the possibility that our model does not account for DM in its entirety, which opens up
the bound to $\Omega h^2 \in [0.06,0.1224]$. For DM direct detection constraints, we apply the current 90\% upper
confidence level cross-sections for spin-independent models with a WIMP mass of $33\gev$, which are given by
$\sigma^{\text{DD-SI}} \leq 7.6 \times 10^{-46} \ \text{cm}^{-2} = 7.6 \times 10^{-10} \ \text{pb}$
\cite{Akerib:2013tjd}. For WIMP masses less or greater than $33\gev$ the direct detection bound is weaker, so this
choice is conservative. 

Furthermore we require agreement with the recently measured Higgs mass, the correct branching ratios for the decays $b
\to s \gamma$ and $B_s \to \mu^+\mu^-$, and agreement with the $\rho$-parameter. The current combined ATLAS and CMS
measurement of the Higgs boson mass is $m_H = (125.09 \pm 0.21\, (\text{stat.}) \pm 0.11\, (\text{sys.})) \ \text{GeV}$
\cite{Aad:2015zhl}. However, these experimental uncertainties are dominated by our much larger theoretical uncertainty,
and consequently we relax our constraint to scenarios with $m_H = (125.09 \pm 1.5)\ \text{GeV}$. We directly apply
limits on the  branching ratios $\text{Br}(b \to s \gamma) = (3.29 \pm 0.19 \pm 0.48) \times 10^{-4}$
\cite{Lees:2012wg} and $\text{Br}(B_s \to \mu^+ \mu^-)$$ = 3.0^{+1.0}_{-0.9} \times 10^{-9}$ \cite{Chatrchyan:2013bka}.

Apart from the fixed experimental constraints, we are free to further modify the parameter space in order to include
some useful features. For example, having light sleptons, especially smuons, is one of these features. This is reasoned
by the fact that light smuons heavily increase the $\Delta a_\mu$ contribution from diagram $(A)$ (see eq.
\ref{eq:loops:A}). Also, having light sleptons grants a suitably higher possibility to explore them during current or
upcoming experimental studies, e.g. at the LHC, due to the comparably clean muonic signals. The corresponding
parameters for the smuons are $m_0$ and $m_2$, which need to be light in order to get light smuons. For actual
parameter choices, see tables \ref{table:Range1}, \ref{table:RangeSmallMu} and \ref{tab:input-parameters}.

Two other useful features are a bino-like LSP (denoted by $\tilde{\chi}$) and a large mass gap between the LSP and the
smuon masses. These characteristics are helpful to provide the correct dark matter relic density while preventing
leptons arising from $\tilde{\mu}^\pm \to \tilde{\chi}\, \mu^\pm$ decays to be soft, which would render them nearly
undetectable at any collider. None of the parameters of our model is directly responsible for these features, so
analysing different scans with different parameter choices is necessary.

As a last point, we have  verified that  benchmarks we consider below do not violate any of the 8 TeV ATLAS and CMS analyses. This is necessary, since one of the scenarios we have found -- the small $\mu$ scenario -- could give rise to light $\tilde{\chi}^0_1$, $\tilde{\chi}^0_2$ and $\tilde{\chi}^\pm_1$ with comparatively low  (few dozen GeV) mass splittings. This region of the parameter space provides distinctive di-lepton or tri-lepton signatures at the LHC which are not observed and which therefore rule out the respective parameter space. To do this verification we have used the chain consisting of \texttt{MadGraph 5.2.2.3}~\cite{Alwall:2014hca} to generate all relevant combinations for chargino-neutralino pair production, \texttt{PYTHIA 6.4}~\cite{Sjostrand:2006za} linked to \texttt{MadGraph} to simulate the parton showering and hadronisation and \texttt{CheckMATE 1.2.1}~\cite{Drees:2013wra} to perform fast detector simulations with \texttt{DELPHES 3.0}~\cite{deFavereau:2013fsa} and event analysis.  Using the same set of cuts as the experimental analyses (either CMS or ATLAS), \texttt{CheckMATE} allowed us  to establish  whether a given point from the parameter space is ruled out or not making use of the data given by the collaborations in their published analyses which are validated in \texttt{CheckMATE}. In particular, we found that tri-lepton signatures explored in Refs. \cite{Aad:2014nua,ATLAS:2013rla} are the most constraining ones for the small $\mu$ region. On the other hand, di-lepton signatures are also worth mentioning, albeit turning out to be less constraining for the parameter space under study.

%% file: 05_1-results_inclusive-scan.tex
\section{Results}
\label{sec:results}

After selecting a certain point in parameter space by choosing all relevant model parameters (cf. section \ref{sec:introduction} and \ref{sec:the-model}), we use \texttt{SoftSUSY 3.5.2} \cite{Allanach:2001kg} to generate the mass spectrum of that point and exclude any point with a Higgs mass out of the bounds chosen in section \ref{sec:constraints}. In case the Higgs mass is in bounds, we use \texttt{micrOMEGAs\_3.6.9.2} \cite{Belanger:2013oya} to compute the relic density as well as the remaining constraints described in section \ref{sec:constraints}.

\subsection{An inclusive scan}
\label{subsec:inclusive}
\begin{figure}[h]
\centering
\includegraphics[width=0.49\textwidth]{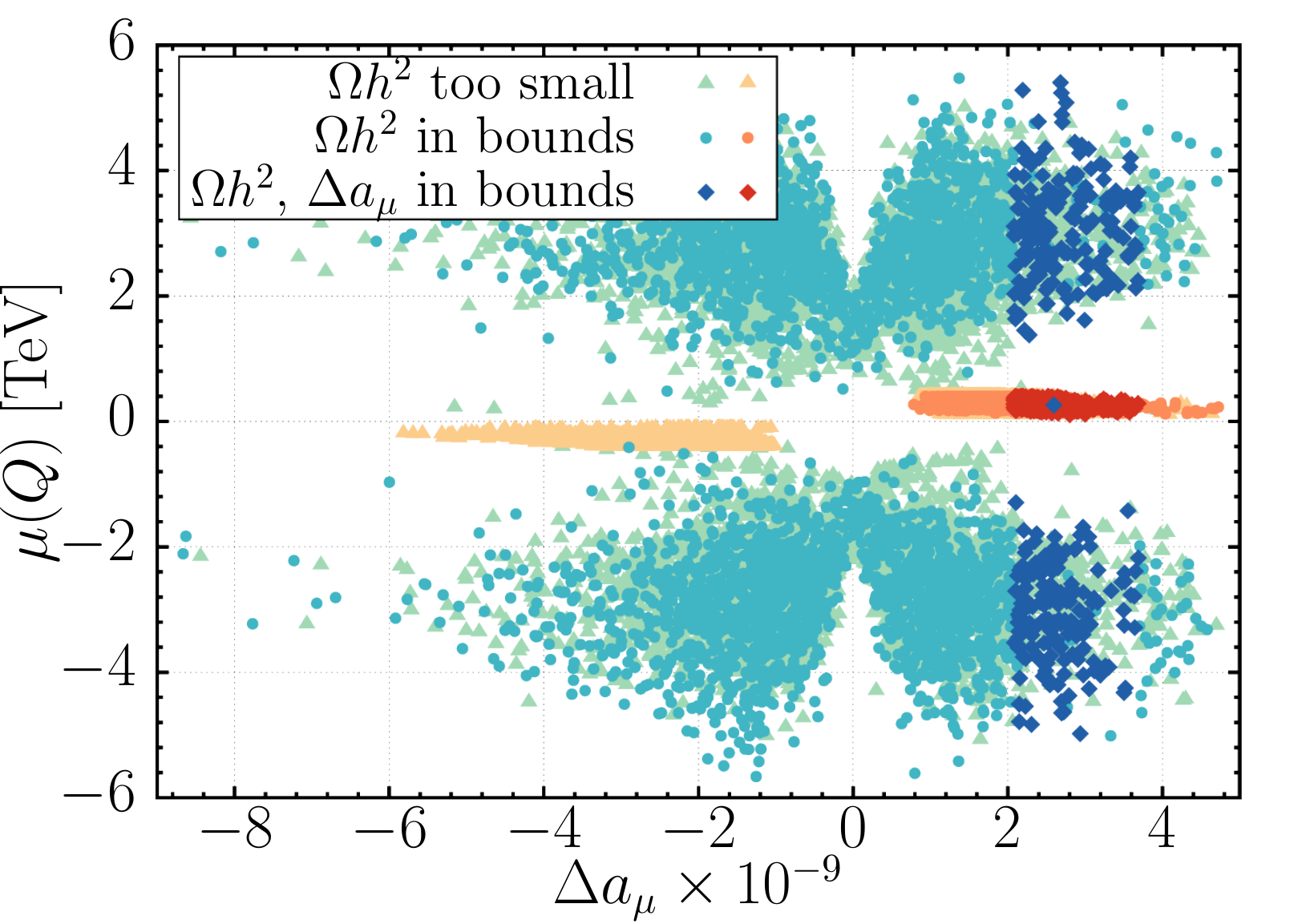}\\
\includegraphics[width=0.49\textwidth]{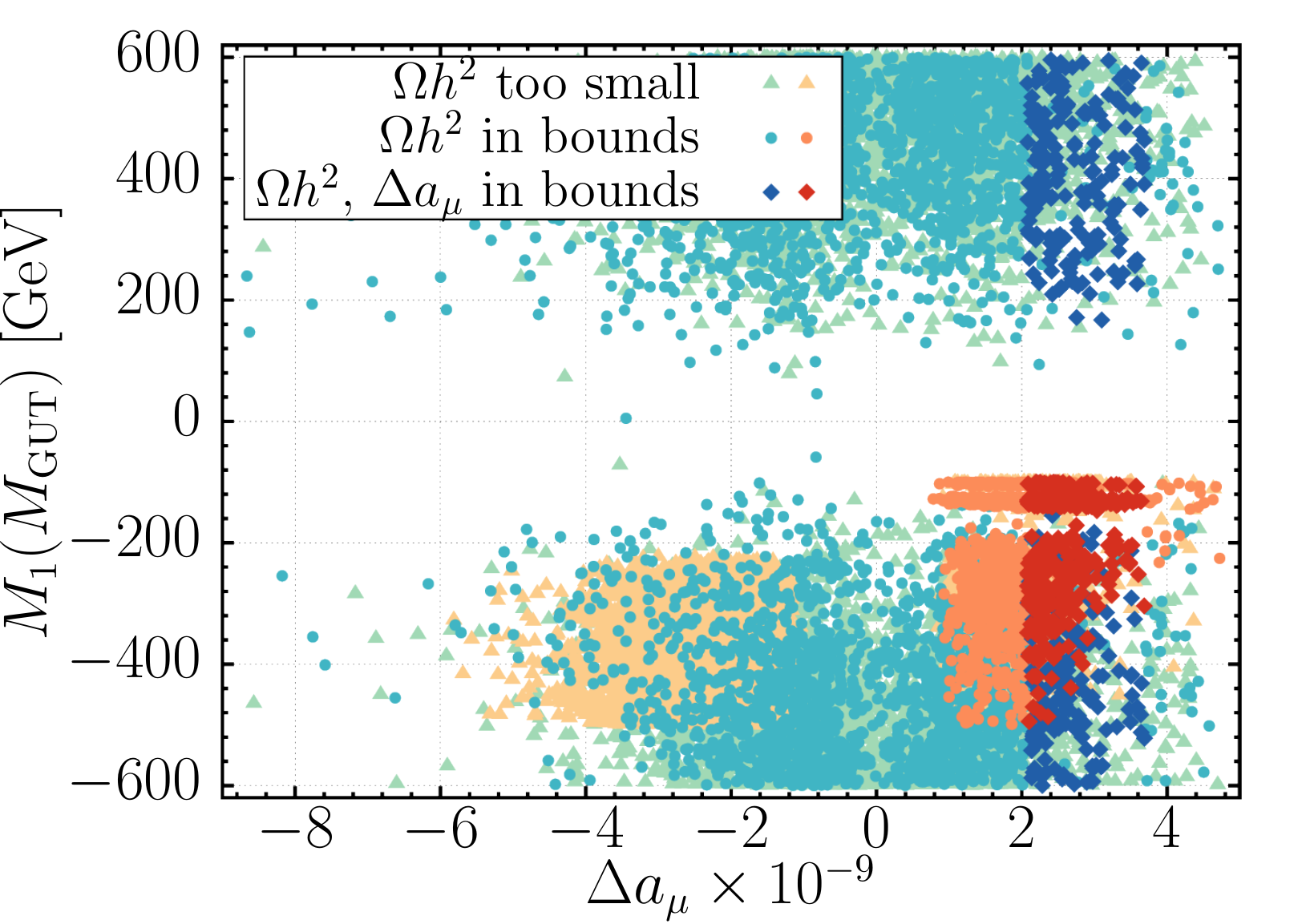}
\includegraphics[width=0.49\textwidth]{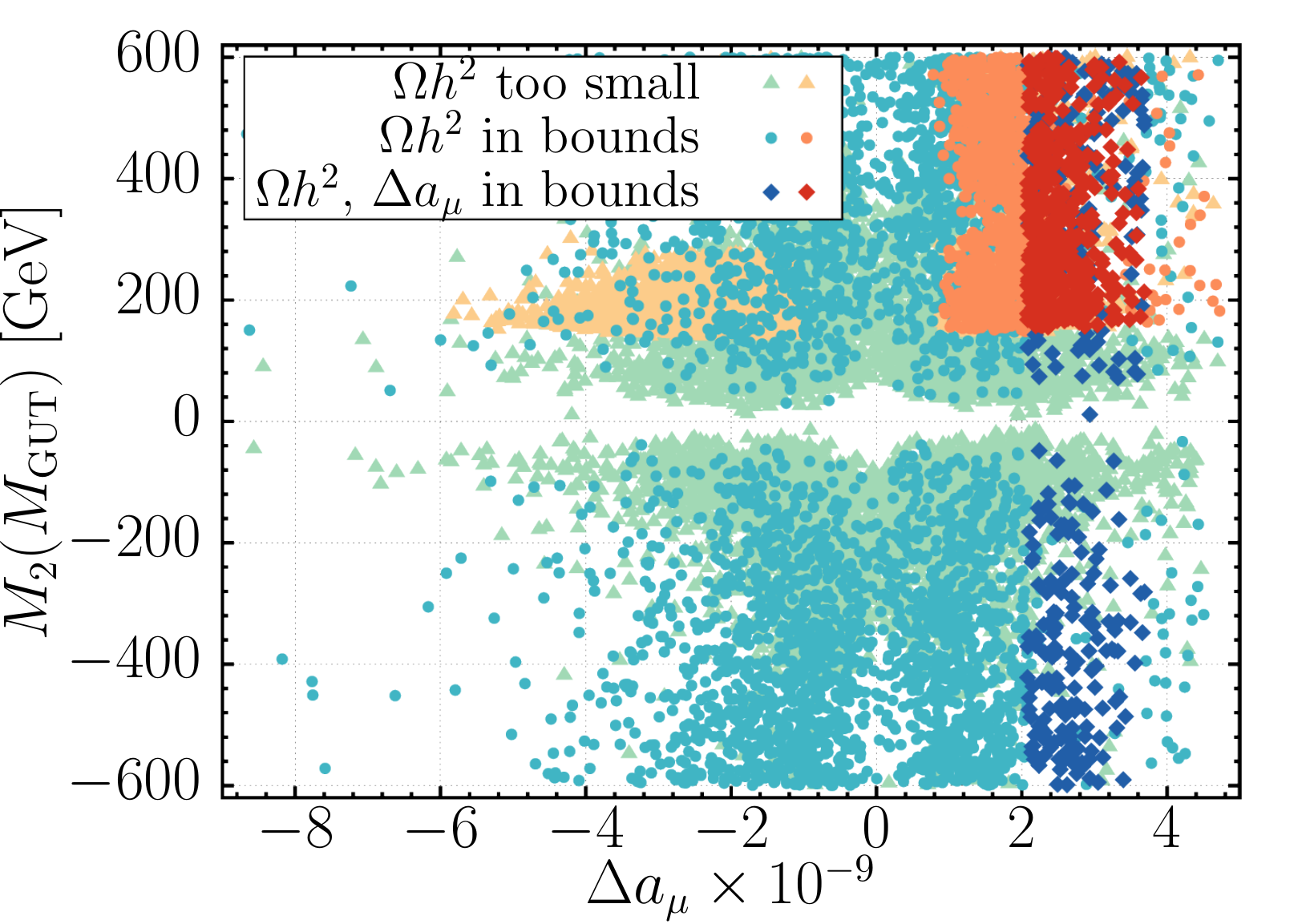}
\caption{Viable scenarios in the $\Delta a_{\mu} \-- \mu$ (top) and $\Delta a_{\mu} \-- M_1$ (bottom-left), $\Delta a_{\mu} \-- M_2$ (bottom-right) planes. Dark blue and red diamonds are scenarios with bino-like DM, whereas the light green and orange triangles and turquoise and salmon circles are scenarios with mainly wino and partially higgsino-like DM. The reddish points correspond to a separate scan around the isolated dark blue point in the top plot at small $\mu$. The input parameters are shown in table \ref{table:RangeSmallMu}.}
\label{fig:gmuon_vs_mu_M1_M2}
\end{figure}

The lack of evidence for strongly interacting superpartners at the LHC puts low scale supersymmetry under pressure. However, while gluinos and squarks of the first two generations need to be heavier than $\sim1.5~\rm{TeV}$, electroweak sector searches are still rather weak. As light supersymmetric particles could be the source for a sizable $\Delta a_{\mu}$ deviation, we investigate scenarios with light smuons and light selectrons in our low scale spectrum that avoid conflict with current experimental exclusion limits. To do this we first preform an inclusive scan on the parameter space varying the GUT scale parameters as shown in table \ref{table:Range1}.
\begin{table}[h]
\centering
\begin{tabular}{|c|rcl|}
\hline
Parameter & \multicolumn{3}{c|}{range} \\
\hline
$\left|A_{\text{tri}}\right|$ & $1$ &--& $3000$ \\
$m_0$, $m_1$, $m_2$ &  $1$ &--& $500$ \\
$m_3$ &  $1$ &--& $3000$ \\
$m_{H_1}$, $m_{H_2}$& $1$ &--& $3000$ \\
\hline
\end{tabular}
\begin{tabular}{|c|rcl|}
\hline
Parameter          & \multicolumn{3}{c|}{range} \\\hline
$\left|M_1\right|$, $\left|M_2\right|$ & $1$ &--& $600$\\
$\left|M_3\right|$ & $1$ &--& $6000$\\ 
$\tan \beta$       & $5$ &--& $50$ \\
$\sgn\(\mu\)$      &     & $\pm 1$ &  \\
\hline
\end{tabular}
\caption{Model parameters at the GUT scale. Dimensionful parameters are in GeV.}
\label{table:Range1}
\end{table}
We allow the $SU(3)_C$ gaugino mass, $M_3$, and the third generation right-handed scalar mass, $m_3$, to acquire large values so the stops may provide a significant contribution to the Higgs mass via loops. In figure \ref{fig:gmuon_vs_mu_M1_M2}, we show viable scenarios in the $\Delta a_{\mu} \-- \mu$ (top), $\Delta a_{\mu} \-- M_1$ (bottom-left) and $\Delta a_{\mu} \-- M_2$ (bottom-right) planes, where the light green and orange triangles have too low relic density, the turquoise and salmon circles have only the relic density in bounds and the dark blue and red diamonds have $\Delta a_\mu$ as well as the relic density in bounds.
%
%
%
It turns out there are two classes of solutions for the correct values of $\Delta a_\mu$, which can be distinguished as a large $\mu$ (the v-shaped bands at $|\mu| \gtrsim 2$ TeV) and a small $\mu$ (the single blue diamond at $\mu \approx 0$ and the red band around it) region. As can be seen, the first class of solutions requires not only a rather large SUSY-preserving mass parameter $\lvert \mu \rvert \gtrsim 2~\rm{TeV}$, but also a soft breaking gaugino mass $\lvert M_1 \rvert  \gtrsim 100~\rm{GeV}$. However, if we relax the relic density requirement, we find solutions from the latter class with small $\mu$ and satisfactory values of $\Delta a_{\mu}$. In particular, the isolated dark blue point in the top of figure \ref{fig:gmuon_vs_mu_M1_M2} at small $\mu$ has $\Delta a_{\mu}=25.96 \times 10^{-10},~\mu = 262.5~\rm{GeV},~M_1 = -475.8~\rm{GeV}~{\rm{and}}~M_2 = 588.9~\rm{GeV}$, and predicts a LSP (bino) with mass $m_{B^0} = 200.1~\rm{GeV}$.

\begin{figure}[H]
\centering
\begin{minipage}{.5\textwidth}
  \centering
  \includegraphics[width=.95\textwidth]{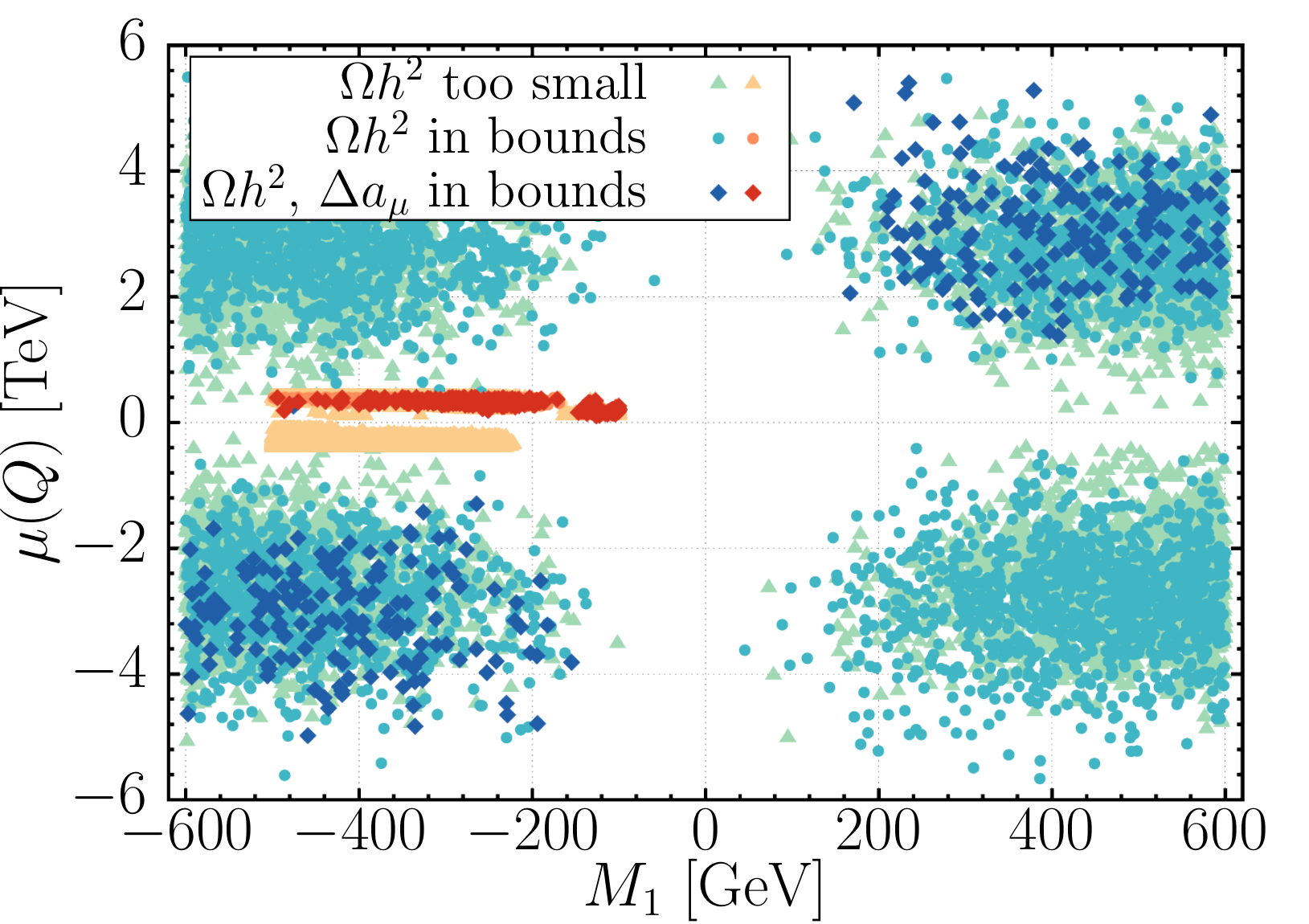}
\end{minipage}%
\begin{minipage}{.5\textwidth}
  \centering
  \includegraphics[width=1.225\textwidth]{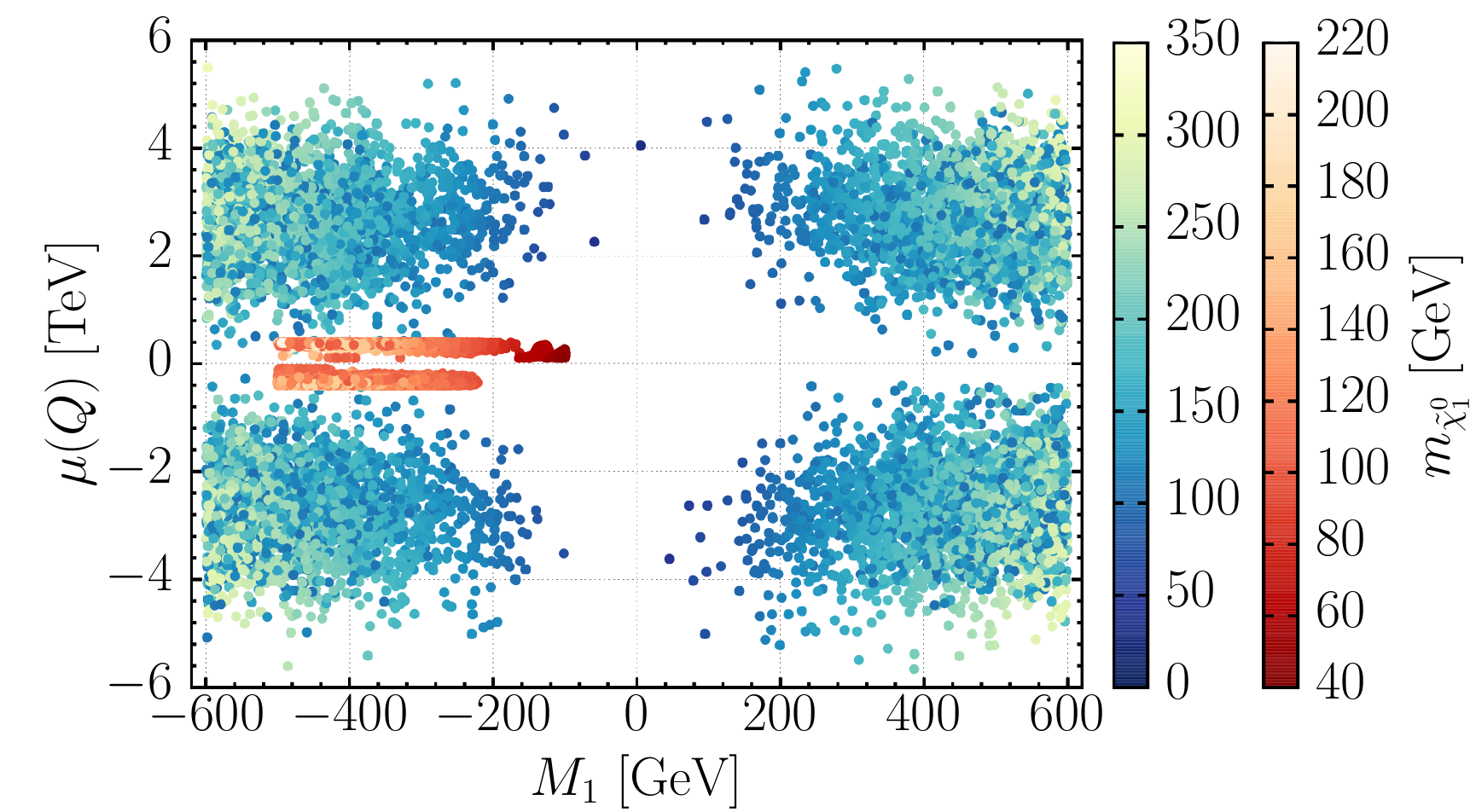}
\end{minipage}
\caption{Correlation between $\mu(Q)$ and $M_1$. The left plot shows points satisfying the given constraints, where the dark blue and red diamonds show points fulfilling both the relic density and $\Delta a_\mu$ constraints, turquoise and salmon circles only have the relic density in bounds, but not $\Delta a_\mu$, and light green and orange triangles refer to points with neither the relic density nor $\Delta a_\mu$ in bounds. The plot on the right shows the same data, but the colour gradients now correspond to the LSP mass $m_{\tilde{\chi}^0_1}$ for the inclusive scan (blue points, see table \ref{table:Range1}) and the small $\mu$ scan (red points, see table \ref{table:RangeSmallMu}), as indicated by the colour bars on the right.}
\label{fig:mu_vs_M1}
\end{figure}

In figure \ref{fig:mu_vs_M1}, we display the correlation between $\mu$, $M_1$ and $m_{\tilde{\chi}^0_1}$, where we have selected only those points where the lightest neutralino wave function is dominated by the bino component. In this figure, we show that for the rare points with light $\mu$, the smallness of the $U(1)_Y$ gaugino mass at the GUT scale ensures that the LSP is predominantly bino via RGE running to the electroweak scale.  
%
%
%

%% file: 05_2-results_small-mu.tex
\subsection{\texorpdfstring{Small $\mu$}{Small mu}}
\label{subsec:small_mu}
\begin{figure}[h] 
  \centering
  \includegraphics[scale=.45]{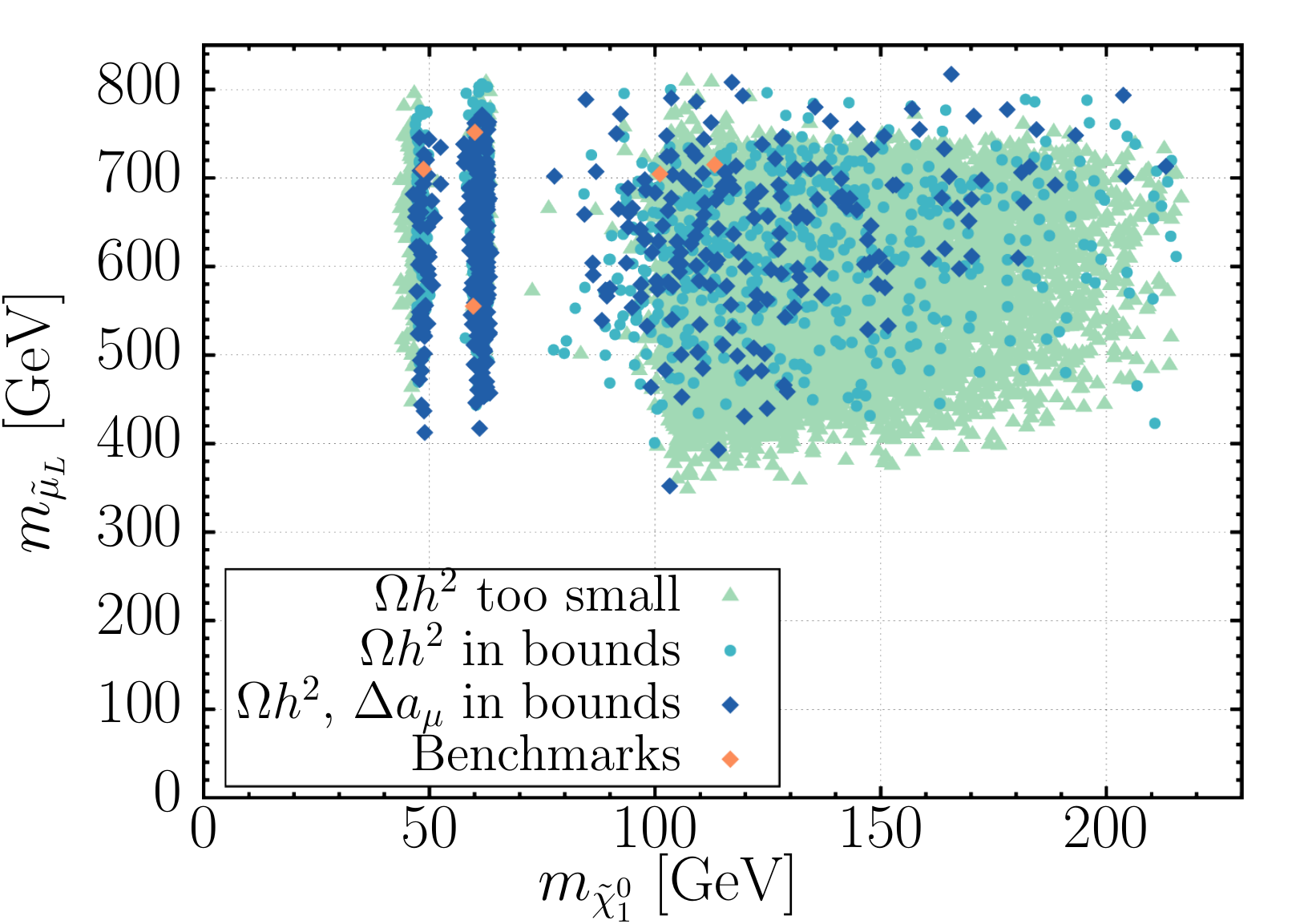}
  \includegraphics[scale=.45]{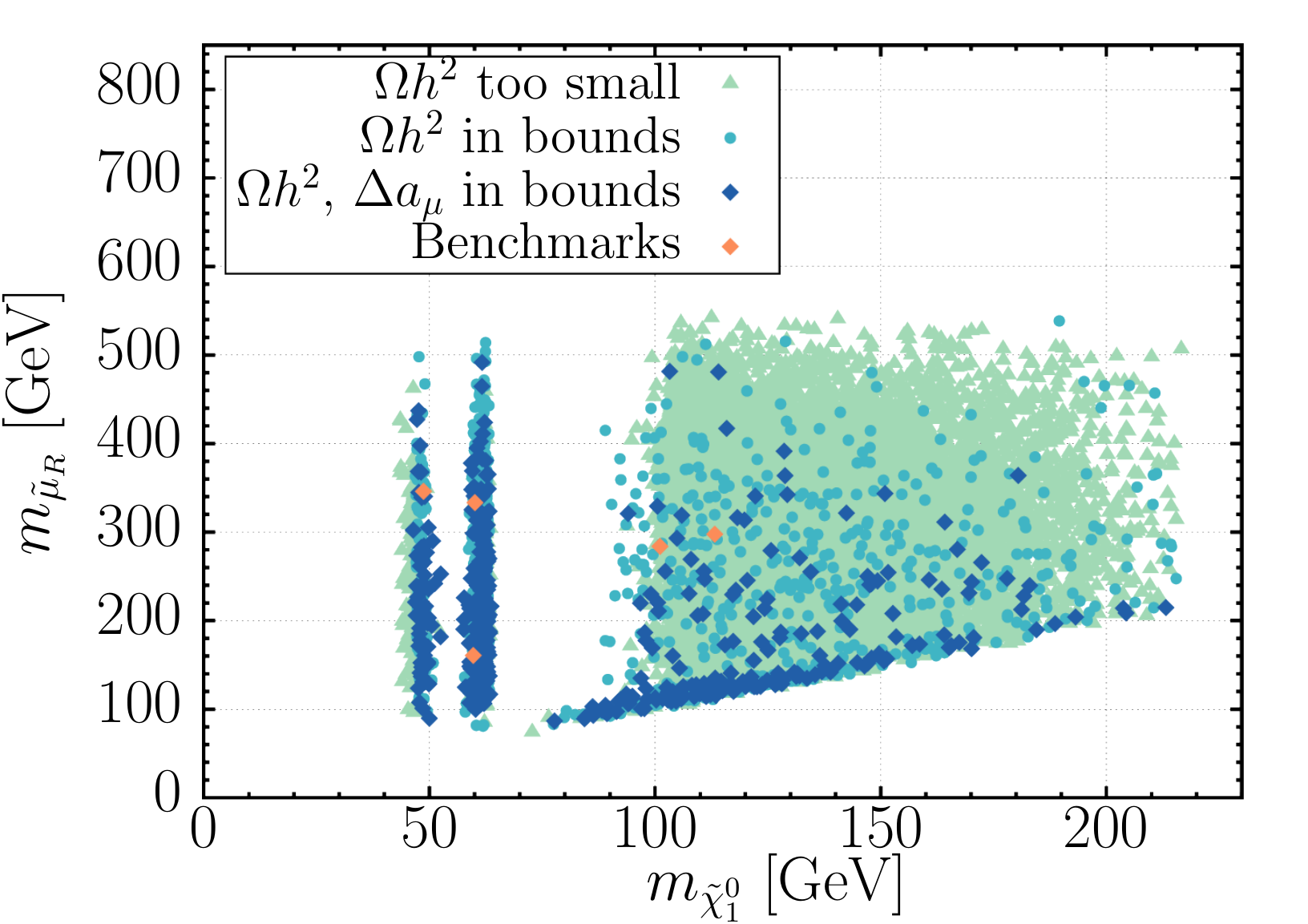}
  \caption{Lightest neutralino mass vs. smuon masses. All dark blue diamonds are bino-like, whereas the light green triangles and turquoise circles are wino-like. The orange pentagons represent the benchmark points defined in table \ref{tab:benchmark_tab_small}.}
  \label{fig:mneut_vs_smuon_small_mu}
\end{figure}

\begin{figure}[H] 
  \centering
  \includegraphics[scale=0.43]{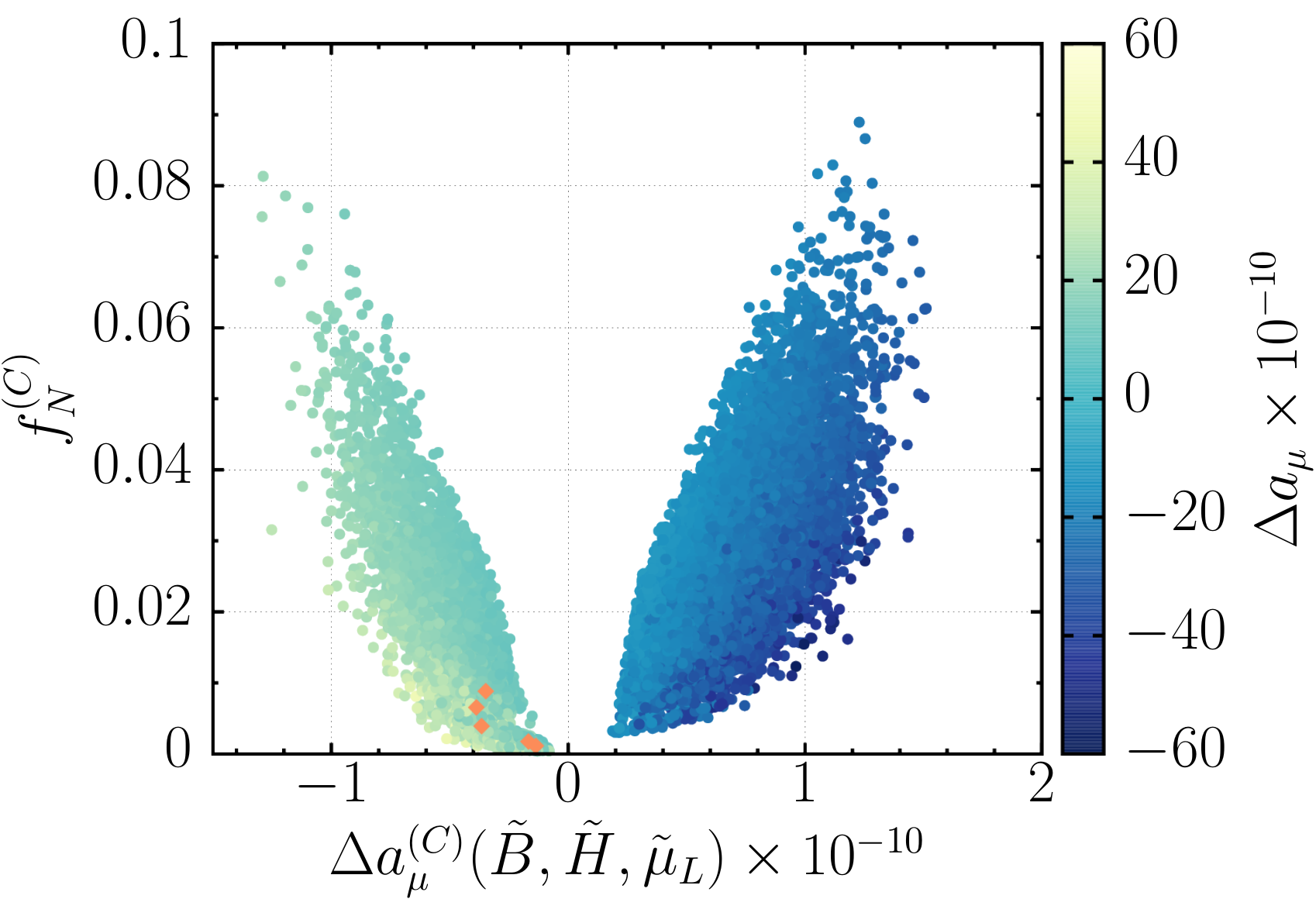}
  \includegraphics[scale=0.43]{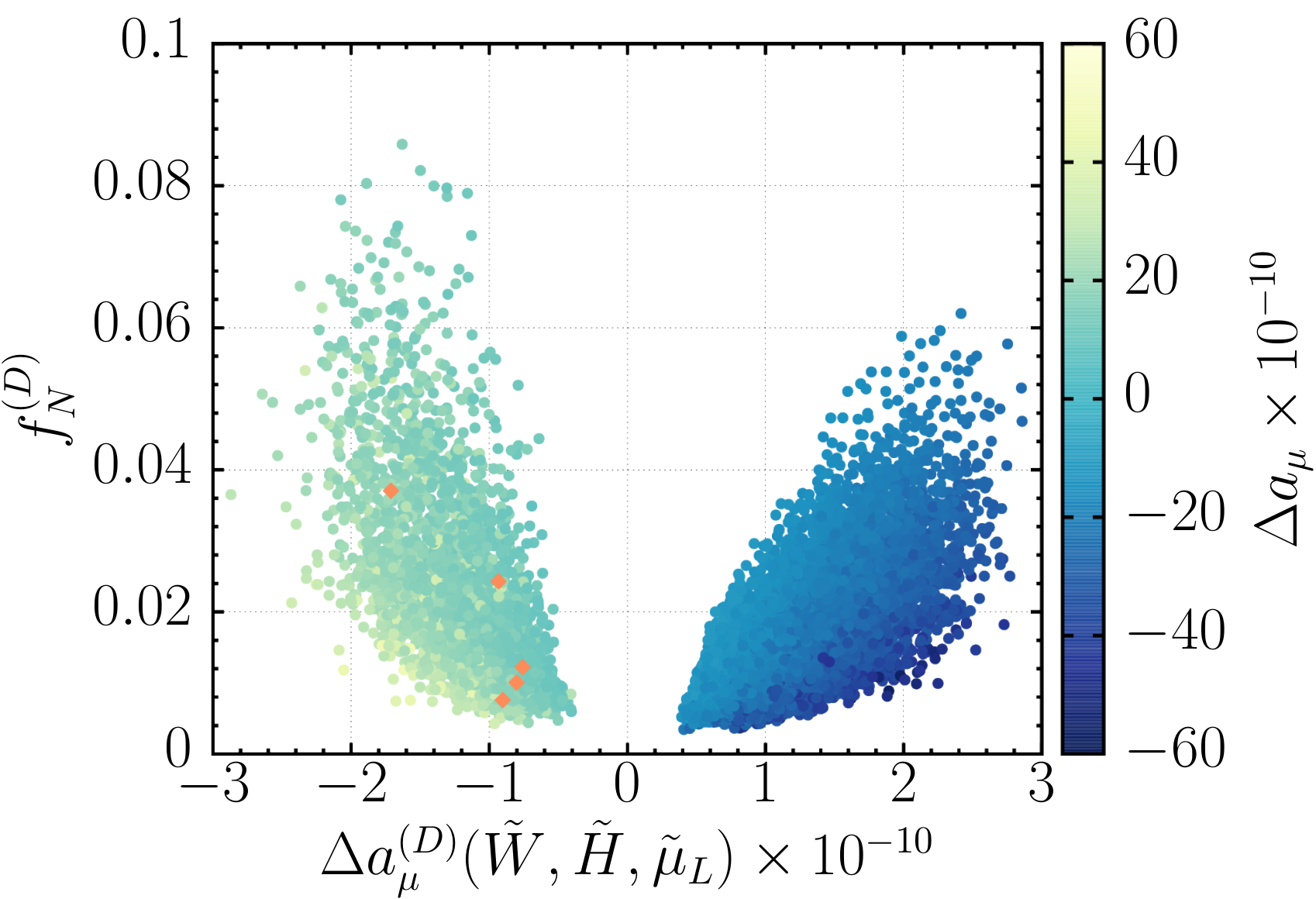}
  \includegraphics[scale=0.43]{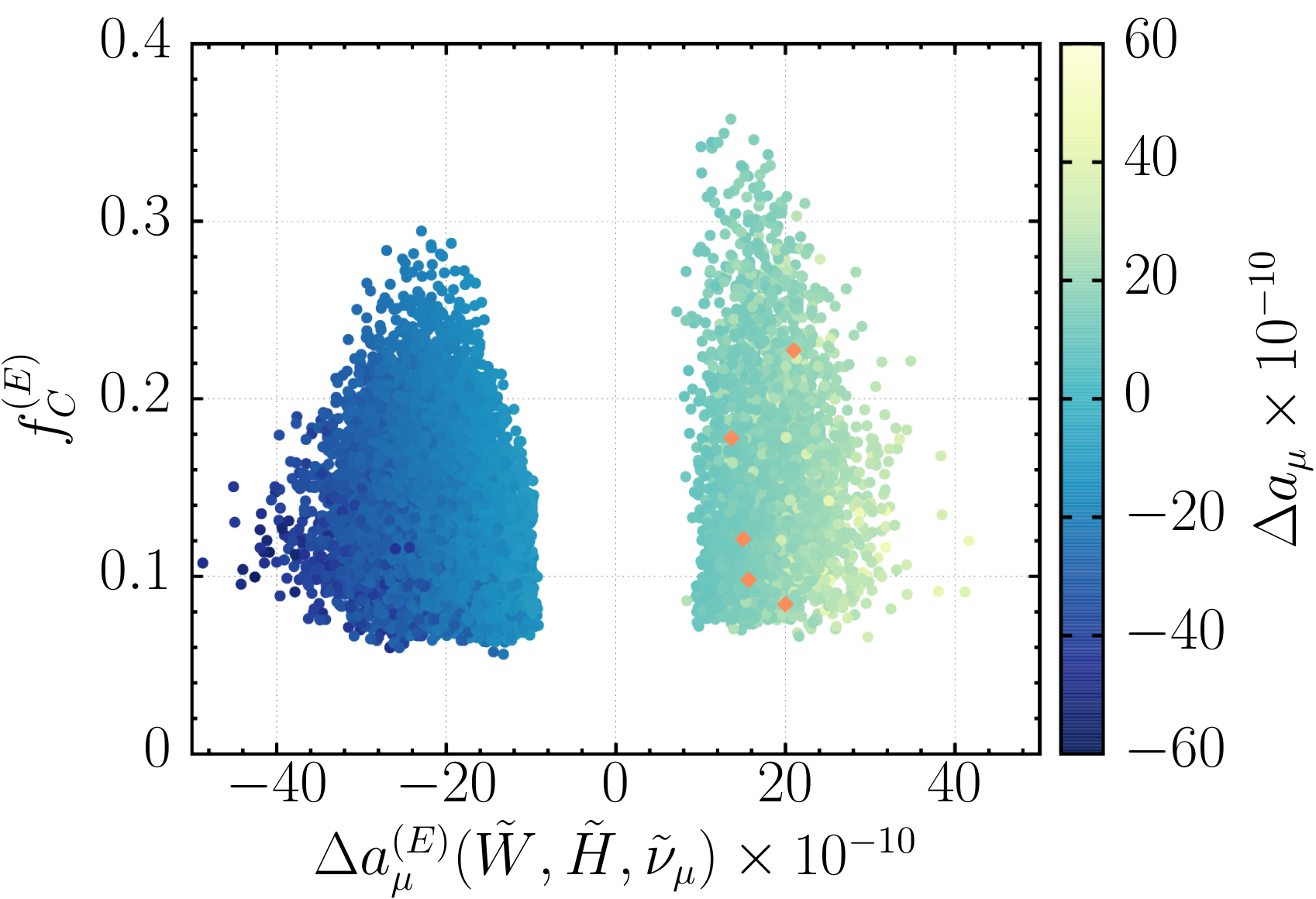}
  \caption{Individual contributions for the $\Delta a^{(i)}_{\mu}$ terms, with $i=\left\{A,B,C,D,E\right\}$ in equations \eqref{eq:loops}, vs. the $f_{N,C}\left( x, y \right)$ functions. The color scale indicates the total value value of $\Delta a_{\mu}$. The orange pentagons represent the benchmark points defined in table \ref{tab:benchmark_tab_small}.}
  \label{fig:f_vs_gmuon}
\end{figure}

As we verified in section \ref{subsec:inclusive}, there are two preferred regions compatible with the correct value for the anomalous magnetic moment of the muon. We  first investigate the small $\mu$ region corresponding to solutions in the vicinity of the isolated band on the top panel of figure \ref{fig:gmuon_vs_mu_M1_M2}. We perform a dedicated scan to generate small $\mu$ and the ranges used for the input parameters at the GUT scale can be found in table~\ref{table:RangeSmallMu}.
\begin{table}[h]
\centering
\begin{tabular}{|c|rcl|}
\hline
Parameter & \multicolumn{3}{c|}{Range} \\
\hline
$A_{\rm{tri}}$ & $-4000$ &--& $-2300$ \\
$m_0$ &  $400$ &--& $700$ \\
$m_1$ &  $300$ &--& $500$ \\
$m_2$ &  $200$ &--& $400$ \\
$m_3$ &  $200$ &--& $2000$ \\
$m_{H_1}$, $m_{H_2}$& $1500$ &--& $2500$ \\
\hline
\end{tabular}
\begin{tabular}{|c|rcl|}
\hline
Parameter & \multicolumn{3}{c|}{Range} \\\hline
$M_1$ & $-500$ &--& $-100$\\
$M_2$ & $100$ &--& $600$\\
$M_3$ & $750$ &--& $1200$\\ 
$\tan \beta$ & $15$ &--& $35$ \\
$\sgn\( \mu \)$&  & $+ 1$ &  \\
&&&\\
\hline
\end{tabular}
    \caption{Theory parameters at the GUT scale. The soft-SUSY breaking parameters are given in GeV.}
    \label{table:RangeSmallMu}
\end{table}
We show in figure \ref{fig:mneut_vs_smuon_small_mu} the results obtained for this scan in the $m_{\tilde{\chi}^0_1}$ vs $m_{\tilde{\mu}_R}$ plane, where only points with positive contribution to the anomalous magnetic moment of the muon are displayed. We observe two clear bands and a bulk region corresponding to distinct regions, where dark matter efficiently annihilates due to different physics processes. In particular, the vertical band with $m_{\tilde{\chi}^0_1}\lesssim 50~{\rm GeV}$ corresponds to LSP annihilation via $Z$ boson resonant decay, whereas the band with $m_{\tilde{\chi}^0_1}\gtrsim 60~{\rm GeV}$ the annihilation into visible SM particles is possible due to Higgs boson exchange.
The lower diagonal band with $m_{\tilde{\chi}^0_1} \sim m_{\tilde{\mu}_R}$ corresponds to the neutralino-smuon co-annihilation region whereas the bulk region on top of this band shows scenarios where dark matter co-annihilates with non-smuon NLSP.
\begin{figure}[h] 
  \centering
  \includegraphics[scale=.45]{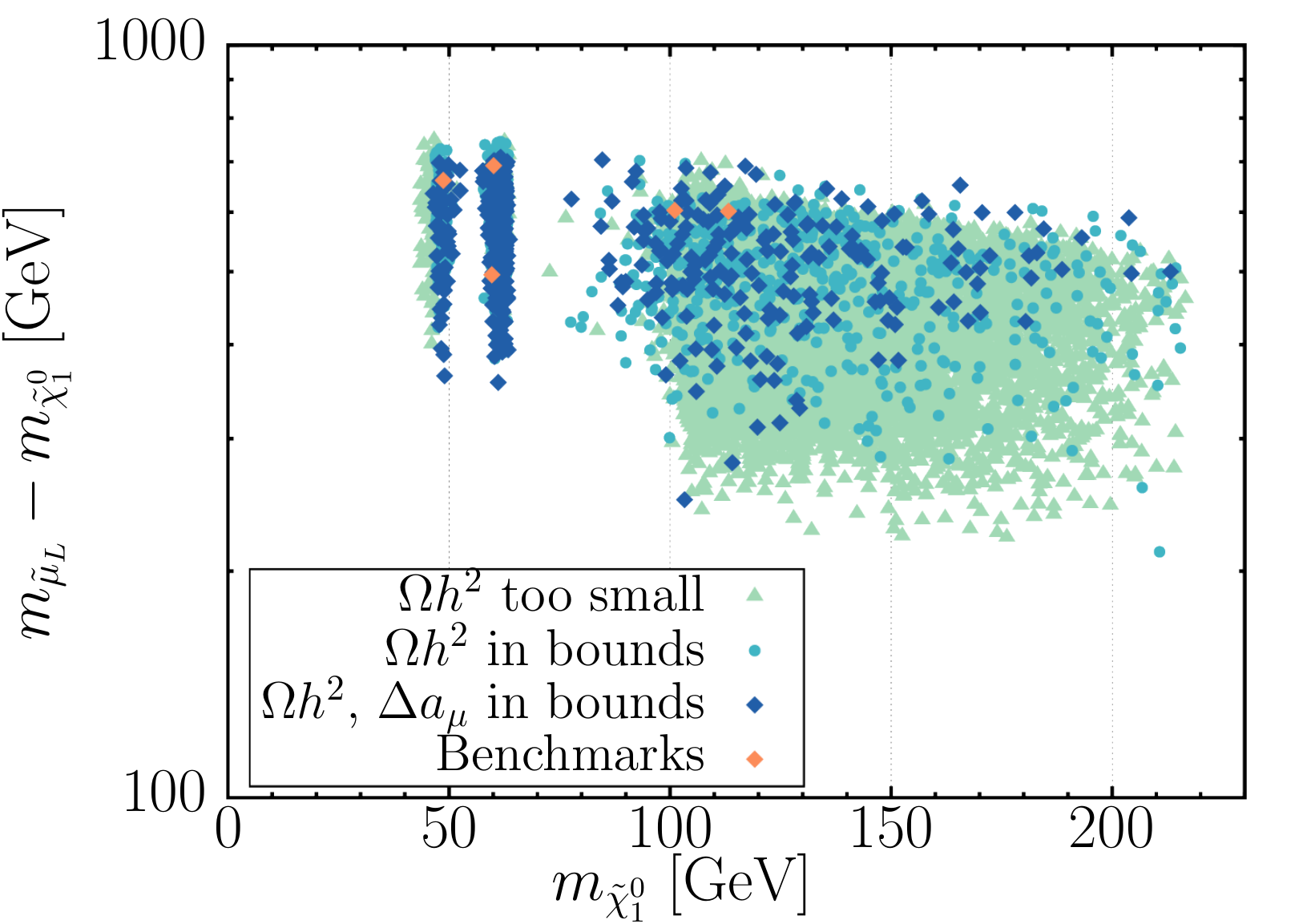}
  \includegraphics[scale=.45]{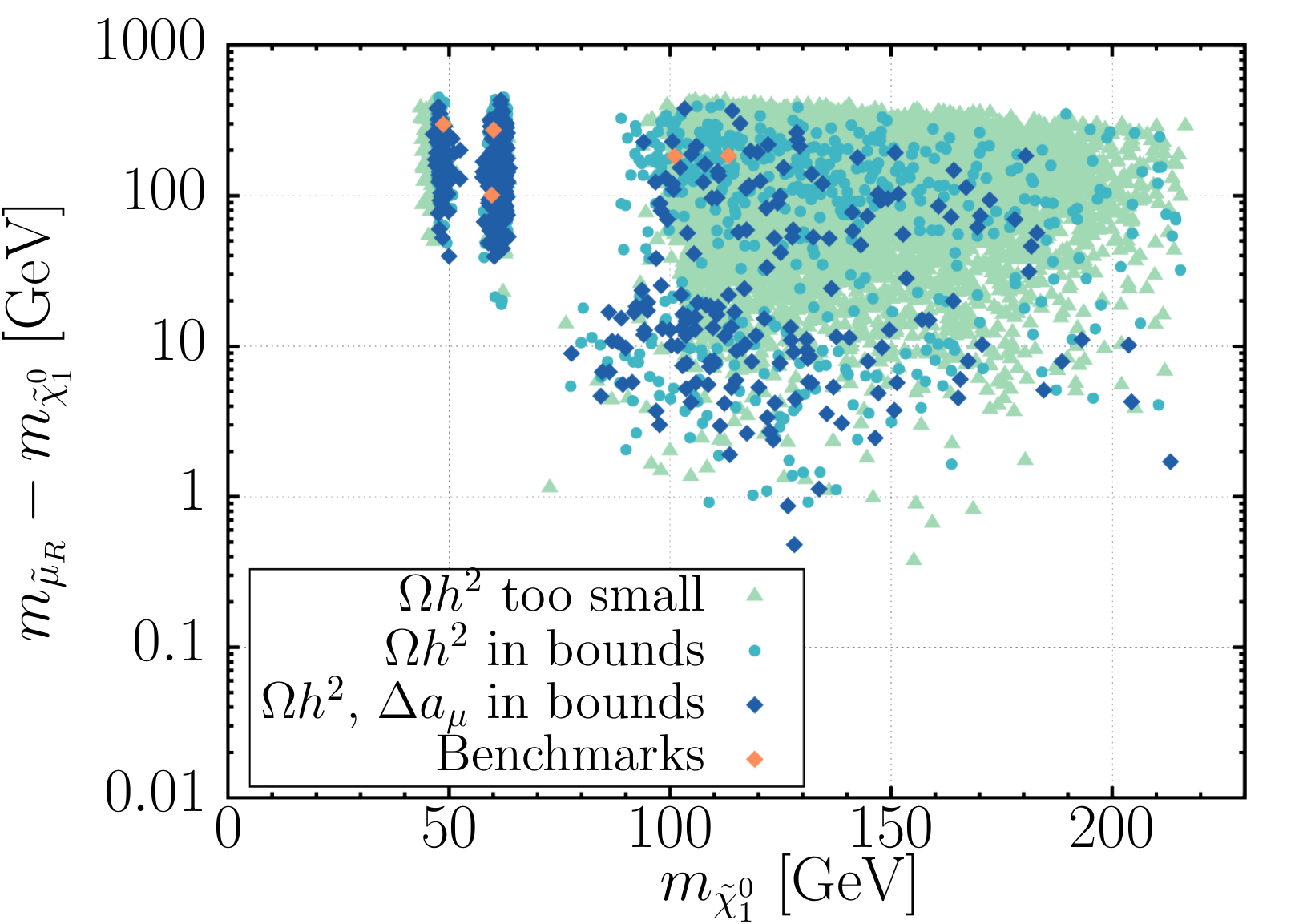}
  \caption{Mass gaps between the smuon and lightest neutralino masses $m_{\tilde{\mu}_{L/R}}$ and $m_{\tilde{\chi}^0_1}$. All dark blue diamonds are bino-like, whereas the light green triangles and turquoise circles are wino-like. The orange pentagons represent the benchmark points defined in table \ref{tab:benchmark_tab_small}.}
  \label{fig:mass_gaps_small_mu}
\end{figure}
In figure \ref{fig:mass_gaps_small_mu}, we show the mass differences for $m_{\tilde{\mu}_{L/R}} - m_{\tilde{\chi}^0_1}$ versus the lightest neutralino mass. While the mass gap for the left handed smuon never deceeds 200 GeV, mass gaps for the right handed smuon can be as small as 1 GeV, thus rendering any muons emerging from smuon decays nearly undetectable. However, the orange benchmark points have both smuon masses $\gtrsim$ 100 GeV, which prevents the muons from smuon decays to be soft. 
For such small values of $\mu$ it may seem that the leading contributions to $\Delta a_\mu$ arise from diagrams $(B),~(C),~(D)$ and $(E)$ in figure \ref{fig:1-loop-amu} as the factor of $\tfrac{1}{\mu}$ in equations \eqref{eq:loops:B} to \eqref{eq:loops:E} becomes large for small $\mu$. However, as we discussed in section \ref{sec:g-2}, the functions $f^{(A,B,C,D)}_{\rm N} \left( x,y \right)$ and $f^{(E)}_{\rm C} \left( x,y \right)$ may also play an important role and should not be disregarded in this analysis. In order to understand which diagrams are indeed relevant for enhancing $\Delta a_{\mu}$ we show in figure \ref{fig:f_vs_gmuon} each individual contribution $\Delta a^{(X)}_{\mu}$ against the corresponding $f_{N,C}\left( x, y \right)$ function and the total $\Delta a_{\mu}$ in the color scale.   
\begin{figure}[H] 
  \centering
  \includegraphics[scale=0.43]{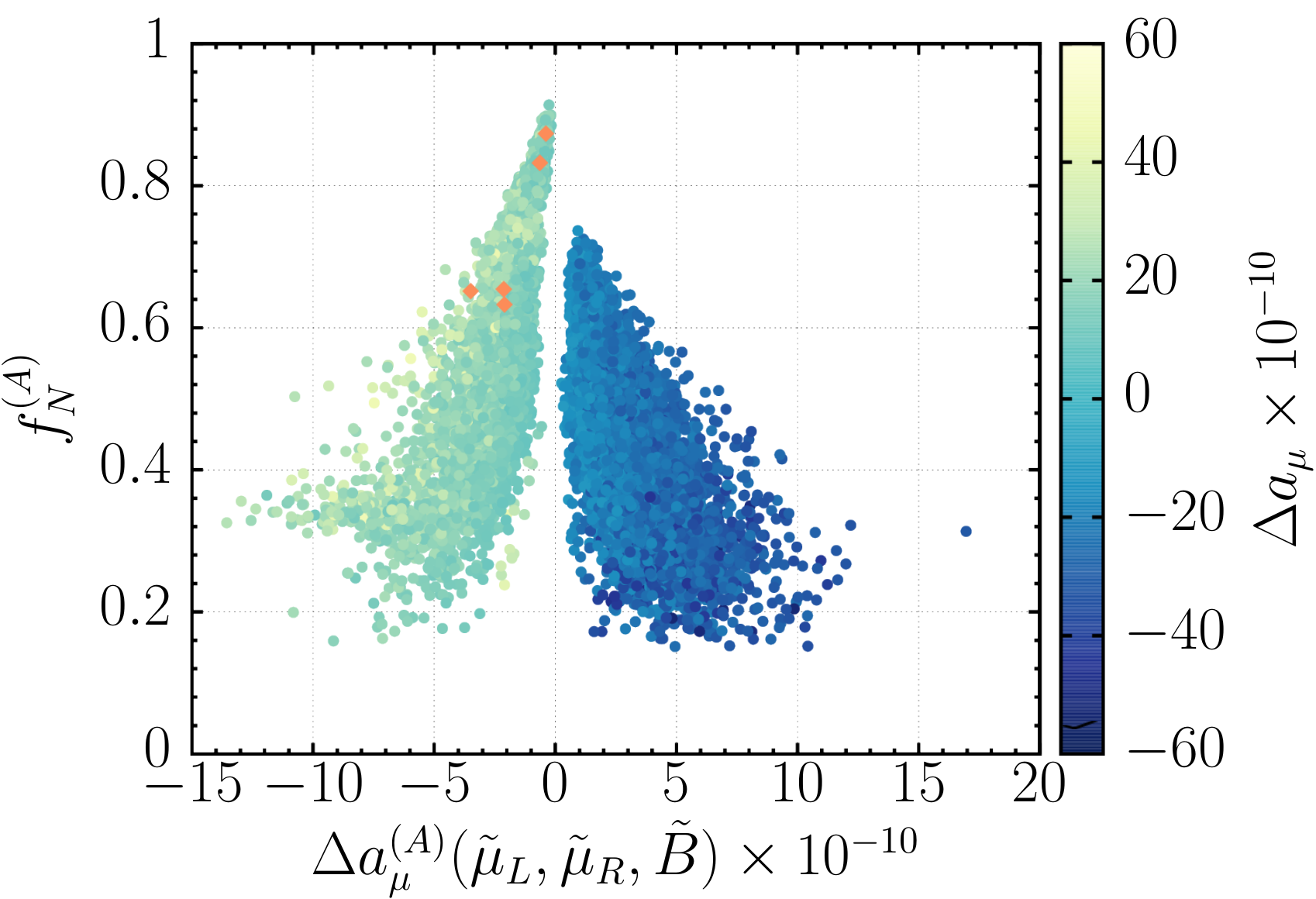}
  \includegraphics[scale=0.43]{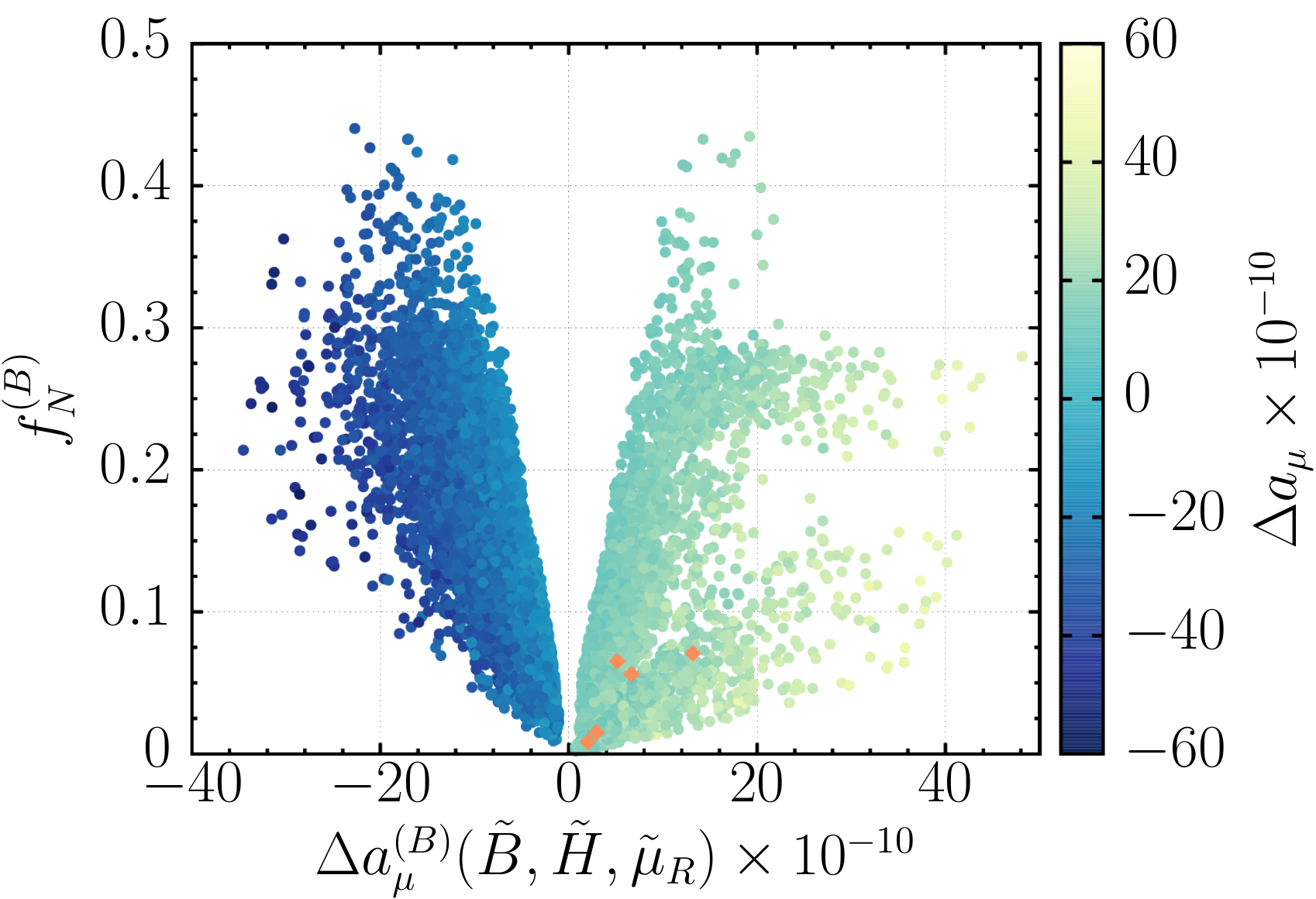}
\end{figure}
The only relevant positive contributions are coming from diagrams $(B)$ and $(E)$. This agrees with equations \eqref{eq:loops:B} and \eqref{eq:loops:E} as in our scan $M_1$ is negative and both $\mu$ and $M_2$ are positive. Furthermore, the leading contributions to $\Delta a_\mu$ are also coming from these two diagrams and the reason for such an enhancement is the dependency on the $f_{\rm N,C} \left( x,y \right)$ functions. In particular, as the right-handed smuon is always lighter than its left-handed counterpart, we have that $f^{(B)}_{{\rm N}}\left(\tfrac{M_{1}^{2}}{m_{\tilde{\mu}_{R}}^{2}},\tfrac{\mu^{2}}{m_{\tilde{\mu}_{R}}^{2}}\right) \gg f^{(C,D)}_{{\rm N}}\left(\tfrac{M_{1,2}^{2}}{m_{\tilde{\mu}_{L}}^{2}},\tfrac{\mu^{2}}{m_{\tilde{\mu}_{L}}^{2}}\right)$, which explains the enhancement of digram $(B)$ and the suppression of the absolute value of $\Delta a^{(X)}_{\mu}$ for diagrams $(C)$ and $(D)$. For the particular case of diagram $(E)$, one could also expect a strong suppression as the muon sneutrino and the left-handed smuon are very close in mass and the $f_{N,C}\left( x, y \right)$ functions share the same asymptotic limits. However this is not what we observe, and if we refer back to the contour plots of figure \ref{fig:contour} and the discussion carried out in section \ref{sec:g-2}, we realise that for the same values of $(x,y)$ we have in general that $f^{(E)}_{C}\left( x, y \right) \gg f^{(A,B,C,D)}_{N}\left( x, y \right)$. Therefore, in diagram $(E)$, it is the function $f^{(E)}_{{\rm C}}\left(\tfrac{M_{2}^{2}}{m_{\tilde{\nu}_{\mu}}^{2}},\tfrac{\mu^{2}}{m_{\tilde{\nu}_{\mu}}^{2}}\right)$ that is responsible for the enhancement of $\Delta a_\mu$, explaining our results. Benchmark points for the small $\mu$ scenario can be found in table \ref{tab:benchmark_tab_small}.
\begin{table}[H]
	\centering
    \resizebox{\textwidth}{!}{%
	\begin{tabular}{cccccccccc}
\toprule
 & Benchmark: & BP1 & BP2 & BP3 & BP4 & BP5 &  \\ 
\midrule
\multirow{12}{*}{\rotatebox{90}{\textsc{Input at GUT scale}}} & $\tan \beta$ & 26.48 & 21.20 & 22.89 & 29.52 & 25.88 \\
 & sgn$(\mu)$       & +       & +       & +       & +       & +       & \\\cline{2-8}
 & $m_0$            & 681.1   & 490.4  & 689.0   & 691.4  & 688.4   & \multirow{10}{*}{\rotatebox{-90}{[GeV]}} \\
 & $m_1$            & 402.0  & 327.5  & 447.0   & 364.4   & 417.9  & \\
 & $m_2$            & 397.4   & 273.0   & 394.2   & 342.2   & 390.7   & \\
 & $m_3$            & 1204.7  & 871.8  & 1085.4  & 987.4  & 1192.3  & \\
 & $M_1$            & -100.1   & -124.1   & -123.8 & -224.9   & -255.1  & \\
 & $M_2$            & 294.9   & 367.5  & 449.9   & 168.6  & 177.9 & \\
 & $M_3$            & 1004.6  & 1085.7  & 1109.8  & 1066.5  & 947.6  & \\
 & $M_{h_1}$      &2204.8  & 2108.4  & 2246.6   & 2127.3  & 2007.2  & \\
 & $M_{h_2}$      & 2385.7   & 2350.9   & 2455.7   & 2330.2 & 2344.7  & \\
 & $A_{\text{tri}}$ & -2839.1 & -2762.5 & -2838.5 & -2764.0 & -3090.0 & \\
\hline
\multirow{24}{*}{\rotatebox{90}{\textsc{Masses}}} & $m_{h_0}$ & 125.2 & 125.2 & 125.2 & 125.1 & 125.1 & \multirow{26}{*}{\rotatebox{-90}{[GeV]}} \\
 & $m_{\tilde{g}}$          & 2220.9  & 2373.5  & 2427.3  & 2349.4  & 2108.5  &  \\
 & $m_{\tilde{q}^{1,2}_L}$  & 2040.6  & 2122.7  & 2220.1  & 2149.1  & 1949.0  &  \\
 & $m_{\tilde{b}_1}$        & 1424.1  & 1537.5  & 1592.3  &  1506.8 & 1234.0  &  \\
 & $m_{\tilde{t}_1}$        & 1120.3  & 1117.4  & 1207.9 & 1184.6 & 962.3  &  \\
 & $m_{\tilde{q}^1_R}$      & 1963.9  & 2086.2  & 2149.9 & 2070.3  & 1872.7 &  \\
 & $m_{\tilde{q}^2_R}$      &1962.9  & 2078.1  & 2136.3 & 2066.3 &  1866.5 &  \\
 & $m_{\tilde{b}_2}$        & 2164.4  & 2108.7  & 2209.8  & 2026.6  & 1984.0 &  \\
 & $m_{\tilde{t}_2}$        & 1488.6  & 1584.3 & 1641.0  & 1561.4  &  1323.4 &  \\
 & $m_{\tilde{e}_L}$        & 710.5   & 555.8   & 752.4   &  705.3  &  715.6  &  \\
 & $m_{\tilde{e}_R}$        & 352.7  & 244.2  & 396.3   &  313.5  &  335.2 &  \\
 & $m_{\tilde{\mu}_L}$      & 710.1   & 555.2   & 751.8   &  704.5  & 714.9  &  \\
 & $m_{\tilde{\mu}_R}$      & 346.1   & 160.7   & 333.5   &  283.9  &  297.6  &  \\
 & $m_{\tilde{\tau}_1}$     & 594.8   & 375.0   & 589.5  &  424.9  &  483.8  &  \\
 & $m_{\tilde{\tau}_2}$     & 1054.1  & 612.5 & 834.6  &  560.1 & 894.9  &  \\
 & $m_{\tilde{\chi}^0_1}$   & -48.58   & -59.58   & -60.00   & -101.0  &  -113.2  &  \\
 & $m_{\tilde{\chi}^0_2}$   & 169.5   & 215.5   & 243.3  & 115.9  &  127.9  &  \\
 & $m_{\tilde{\chi}^0_3}$   & -228.2 & -265.1 & -277.4 &  -350.7 & -411.9 &  \\
 & $m_{\tilde{\chi}^0_4}$   & 287.7  & 337.3  & 391.5 & 357.2 & 416.9  &  \\
 & $m_{\tilde{\chi}^\pm_1}$ & 171.3   & 217.3   & 245.0  &  116.3  &  128.2  &  \\
 & $m_{\tilde{\chi}^\pm_2}$ & 287.4  & 336.9  & 390.8  &  360.4 & 419.9 &  \\
 & $m_{\tilde{\nu}^e_L}$    & 705.8   & 549.9   & 747.9  &  700.5  &  711.0  &  \\
 & $m_{\tilde{\nu}^\mu_L}$  & 705.5   & 549.4   & 747.5   &  704.5  &  710.4  &  \\
 & $m_{\tilde{\nu}^\tau_L}$ & 589.5   & 367.5   & 584.5  &  421.6  &  478.1  &  \\\cline{1-7}
 & $Q$                      & 1293.4 & 1337.0  & 1409.0  & 1360.4  & 1143.6  &  \\
 & $\mu(Q)$                 & 212.3  & 250.5  & 263.2  & 335.2  & 397.9  &  \\
\hline
\multirow{5}{*}{\rotatebox{90}{\textsc{Constraints}}} & Br$(b \to s \gamma)$      & $2.89 \times 10^{-4}$  & $2.91 \times 10^{-4}$  & $2.91 \times 10^{-4}$  & $3.25 \times 10^{-4}$  & $ 3.25 \times 10^{-4}$  &  \\
 & Br$(B_s \to \mu^+ \mu^-)$ & $2.69 \times 10^{-9}$  & $2.97 \times 10^{-9}$  & $2.97 \times 10^{-9}$  & $ 3.06 \times 10^{-9}$  & $3.11\times 10^{-9}$  &  \\
 & $\sigma^{\text{DD SI}}$   & $1.31 \times 10^{-11}$ & $1.28 \times 10^{-11}$ & $1.18 \times 10^{-11}$ & $ 2.42 \times 10^{-11}$ & $1.06 \times 10^{-11}$ & [pb]  \\
 & $\Omega h^2$              & $1.05 \times 10^{-1}$  & $1.25 \times 10^{-1}$  & $1.23 \times 10^{-1}$  & $8.32 \times 10^{-2}$  & $ 8.47 \times 10^{-2}$  &  \\
 & $\Delta a_\mu$            & $1.37 \times 10^{-9}$  & $2.28 \times 10^{-9}$  & $1.30 \times 10^{-9}$  & $1.99 \times 10^{-9}$  & $1.52 \times 10^{-9}$  &  \\
\bottomrule
    \end{tabular}}
    \caption{Input and Output parameters for the benchmark points with the most accurate $\Delta a_\mu$ and $\Omega h^2$ in the case of small $\mu(Q)$ and all other constraints being fulfilled. $\tilde{q}^{i}$ labels the $i$-th generation of squarks.}
    \label{tab:benchmark_tab_small}
\end{table}

%% file: 05_4-results_large-mu2.tex
\subsection{\texorpdfstring{Large $\mu$}{Large mu}}
\label{subsec:large_mu}
The other class of solutions that provides the full contribution to the anomalous magnetic moment of the muon requires $\lvert \mu \rvert \gtrsim 2~{\rm TeV}$. In order to study this region in detail we perform an enhanced scan on the parameter space around the points in figure \ref{fig:gmuon_vs_mu_M1_M2} that better approach the value of $\Delta a_{\mu}$ as given in \eqref{eq:Damu_3}. The new scenarios were generated with the GUT scale parameters as in table \ref{tab:input-parameters}.
\begin{table}[h]
\centering
\begin{tabular}{|c|rcl|}
\hline
Parameter & \multicolumn{3}{c|}{Range} \\
\hline
$A_{\rm{tri}}$ & $-3000$ &--& $0$ \\
$m_0$ &  $100$ &--& $300$ \\
$m_1$ &  $500$ &--& $1500$ \\
$m_2$ &  $100$ &--& $400$ \\
$m_3$ &  $1000$ &--& $2000$ \\
$m_{H_1}$, $m_{H_2}$& $100$ &--& $3000$ \\
\hline
\end{tabular}
\begin{tabular}{|c|rcl|}
\hline
Parameter & \multicolumn{3}{c|}{Range} \\\hline
$M_1$ & $-1000$ &--& $1000$\\
$M_2$ & $-2000$ &--& $2000$\\
$M_3$ & $2000$ &--& $3000$\\ 
$\tan \beta$ & $5$ &--& $50$ \\
$\sgn\( \mu \)$&  & $\pm 1$ &  \\
&&&\\
\hline
\end{tabular}
    \caption{Theory parameters at the GUT scale. The soft-SUSY breaking parameters are given in GeV.}
    \label{tab:input-parameters}
\end{table}

\begin{figure}[H] 
  \centering
  \includegraphics[scale=0.43]{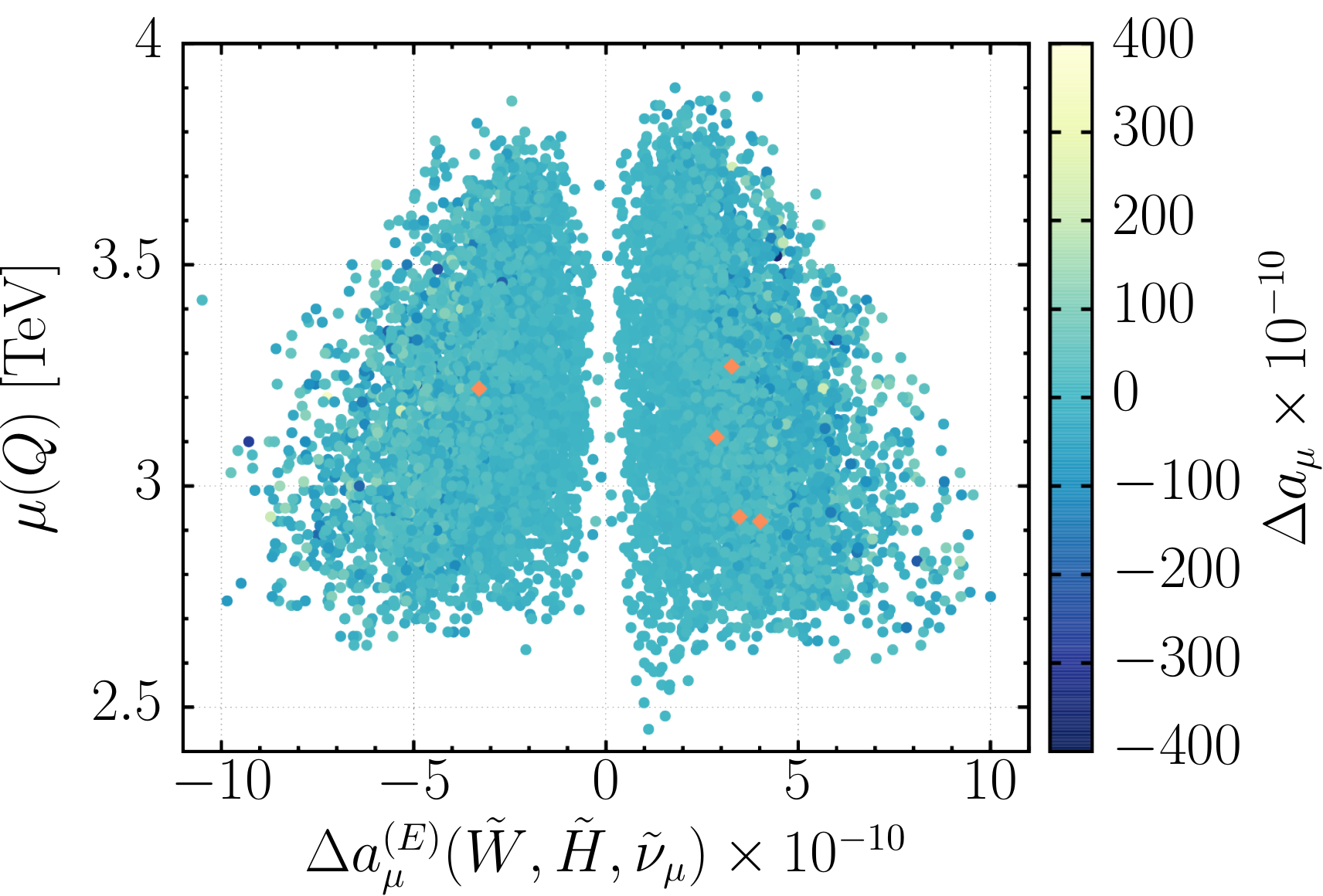}
  \caption{Individual contributions for the $\Delta a^{(i)}_{\mu}$ terms, with $i=\left\{A,B,C,D,E\right\}$ in equations \eqref{eq:loops}, vs. $\mu(Q)$. The color scale indicates the total value value of $\Delta a_{\mu}$, while the black bars in the top left panel show the 1$\sigma$ bound of $\Delta a_\mu$. The orange pentagons represent the benchmark points defined in table \ref{tab:benchmark_tab1}.}
  \label{fig:diag_vs_mu}
\end{figure}

Analogue to figure \ref{fig:f_vs_gmuon}, we first investigate which loop diagrams from equation \ref{eq:loops} contribute most to $\Delta a_\mu$. This is shown in figure \ref{fig:diag_vs_mu}.
\begin{figure}[H] 
  \centering
  \includegraphics[scale=0.43]{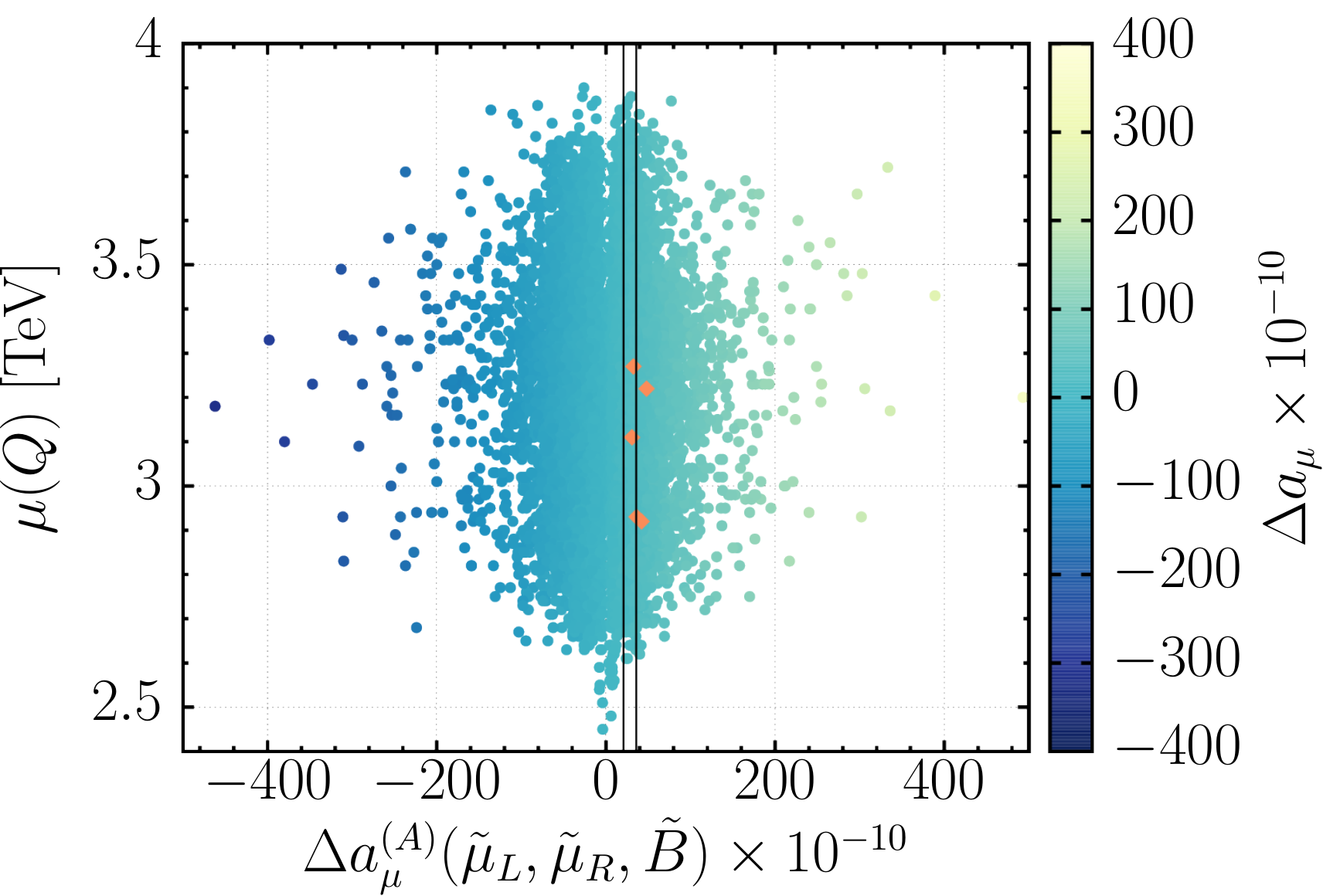}
  \includegraphics[scale=0.43]{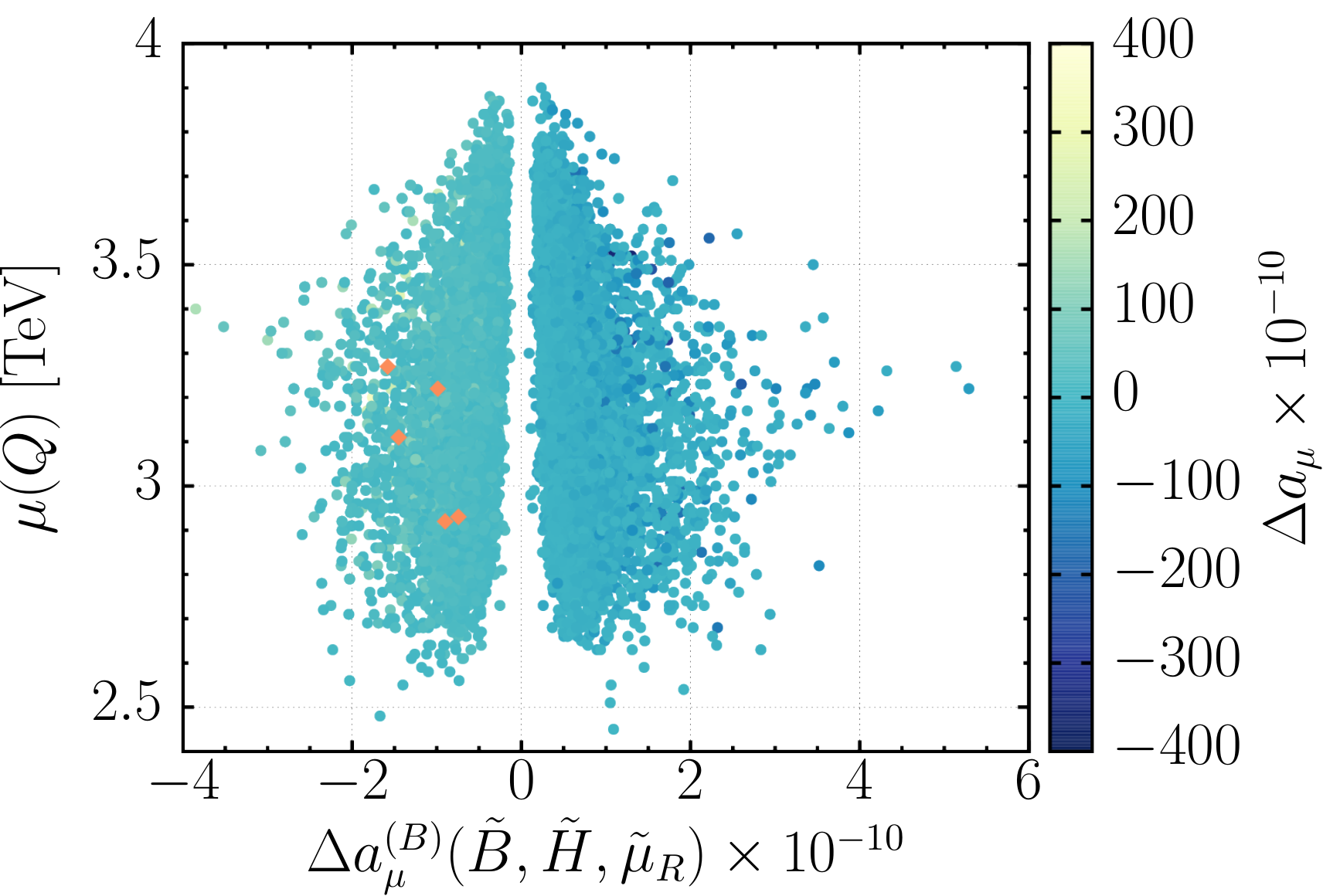}\\
  \includegraphics[scale=0.43]{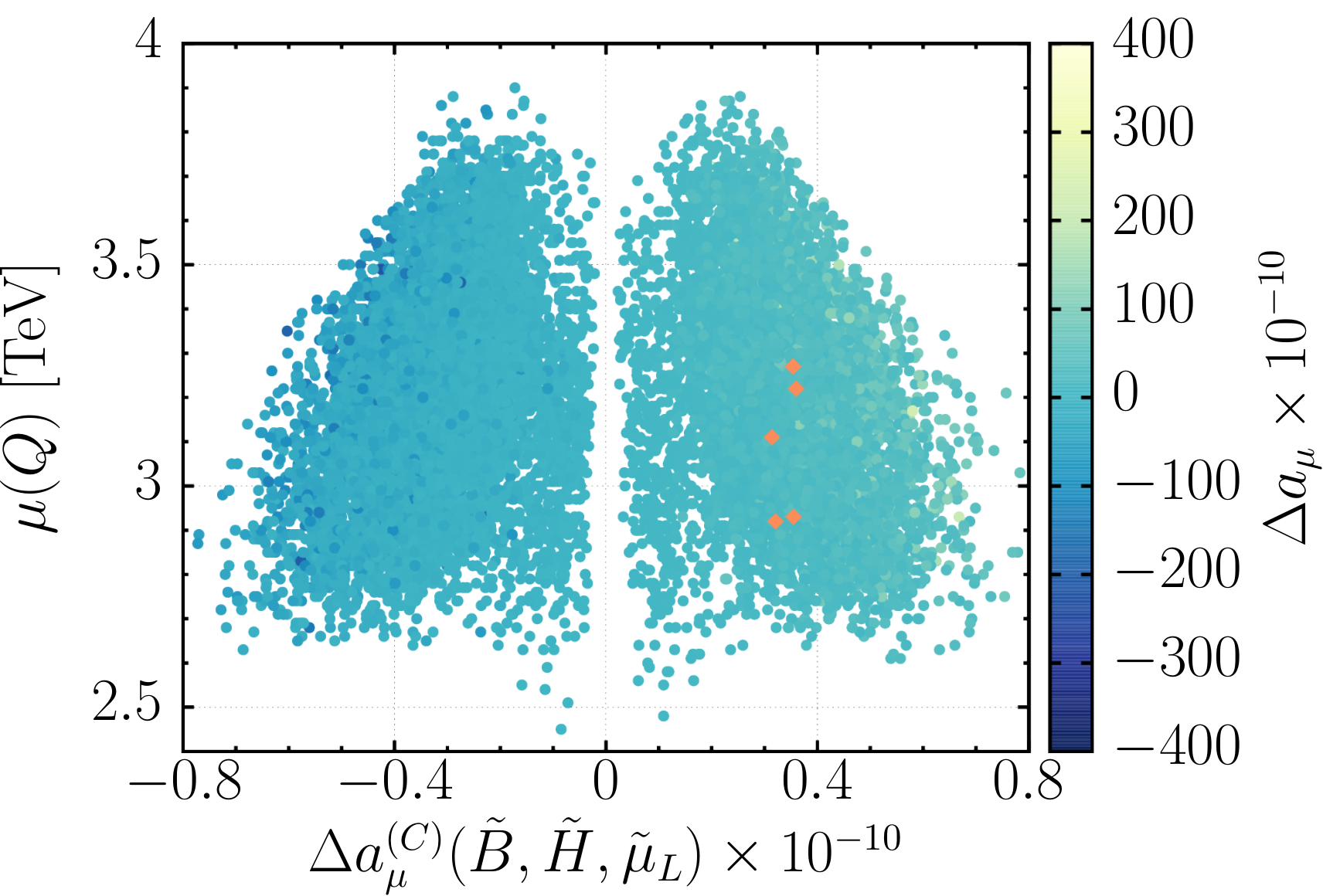}
  \includegraphics[scale=0.43]{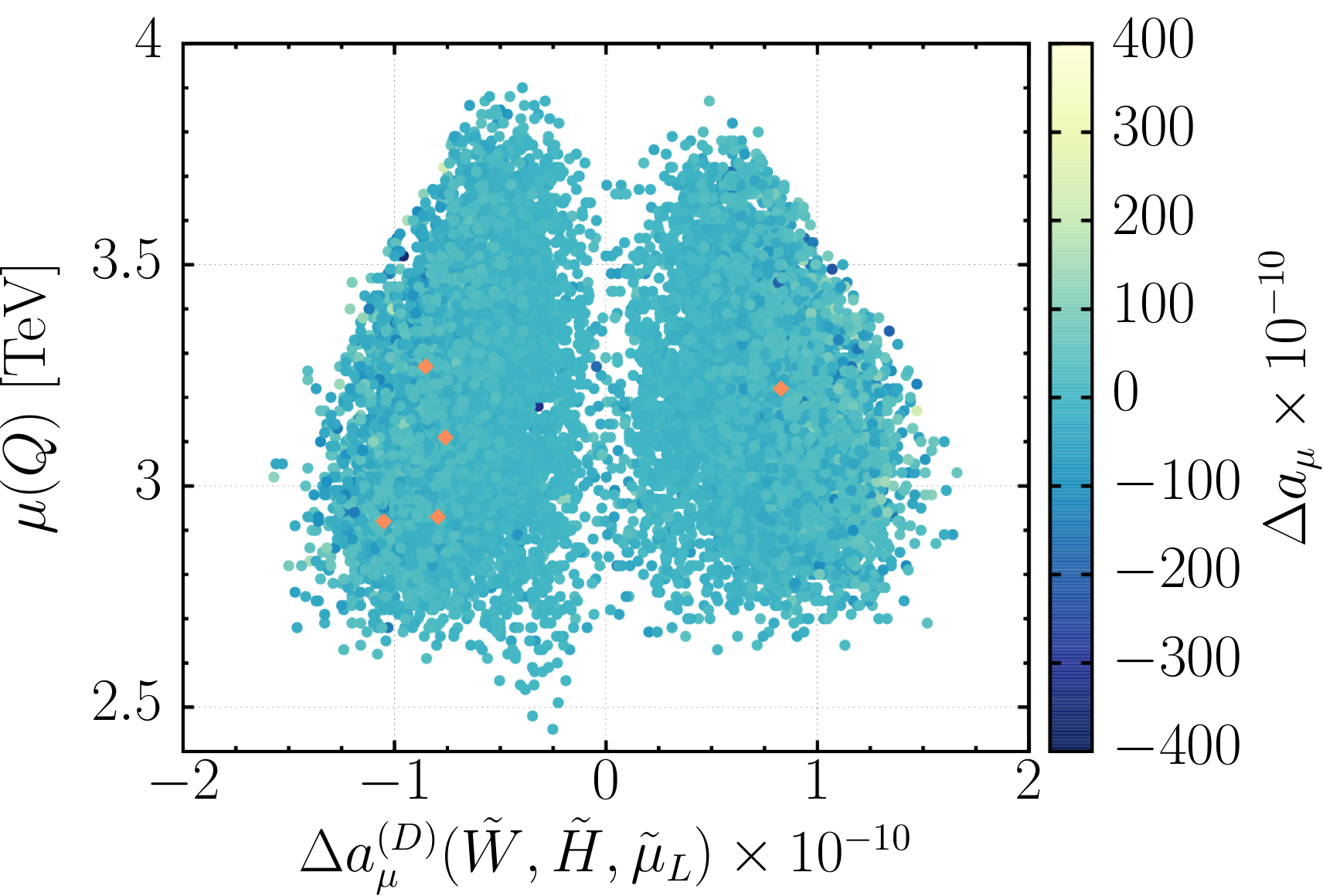}
\end{figure}

\begin{figure}[h] 
  \centering
  \includegraphics[width=\linewidth]{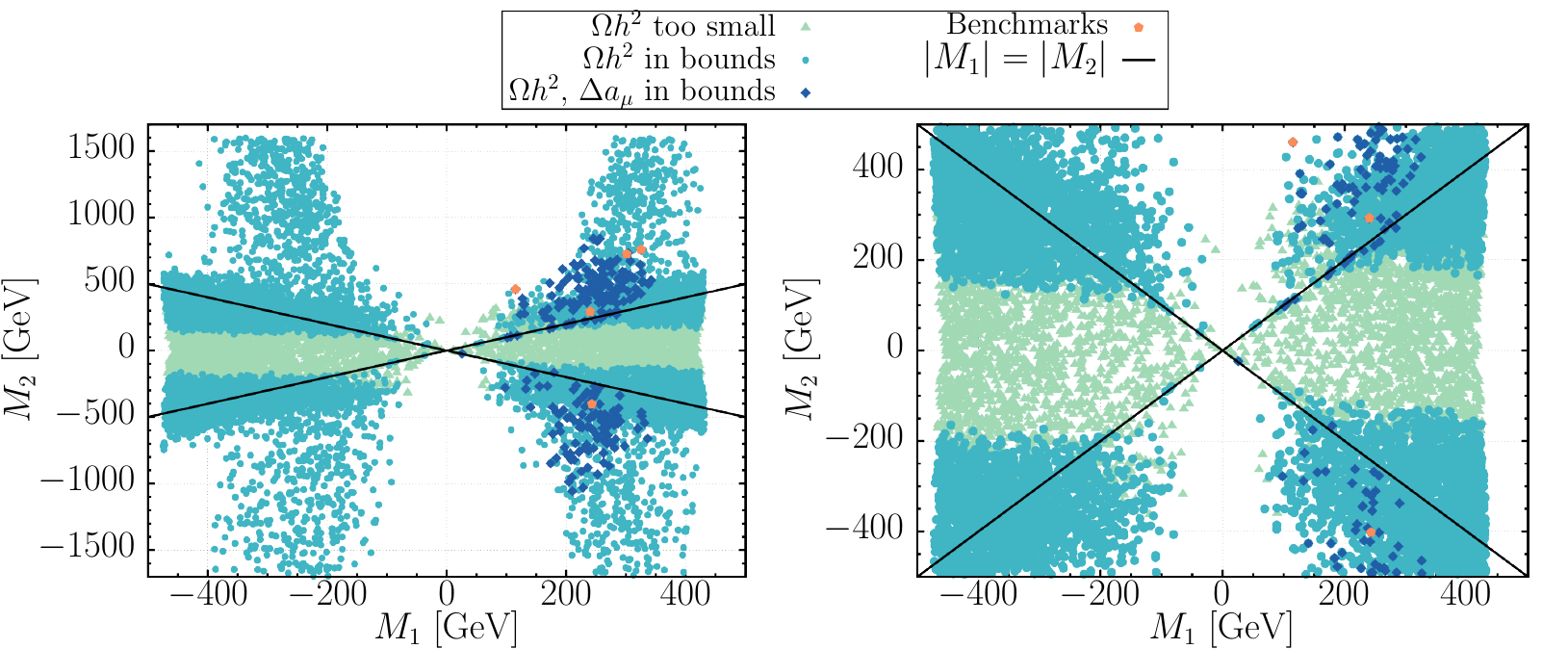}
  \caption{$M_1(Q)$ vs. $M_2(Q)$ (left) and a smaller excerpt of it (right). All dark blue diamonds are bino-like, whereas the light green triangles and turquoise circles are wino-like. The orange pentagons represent the benchmark points defined in table \ref{tab:benchmark_tab1}.}
  \label{fig:M1_vs_M2}
\end{figure}

\begin{figure}[h] 
  \centering
  \includegraphics[scale=.45]{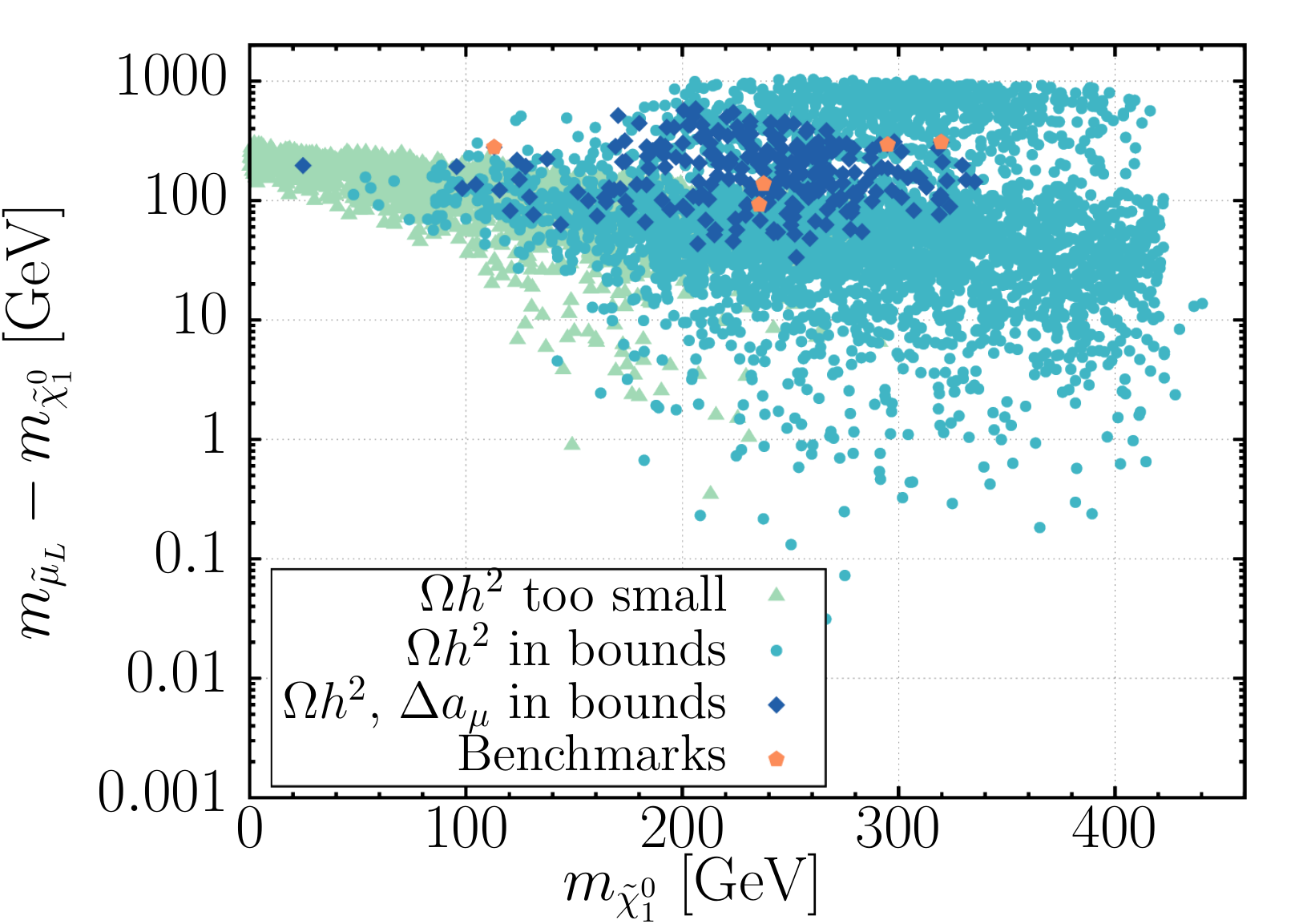}
  \includegraphics[scale=.45]{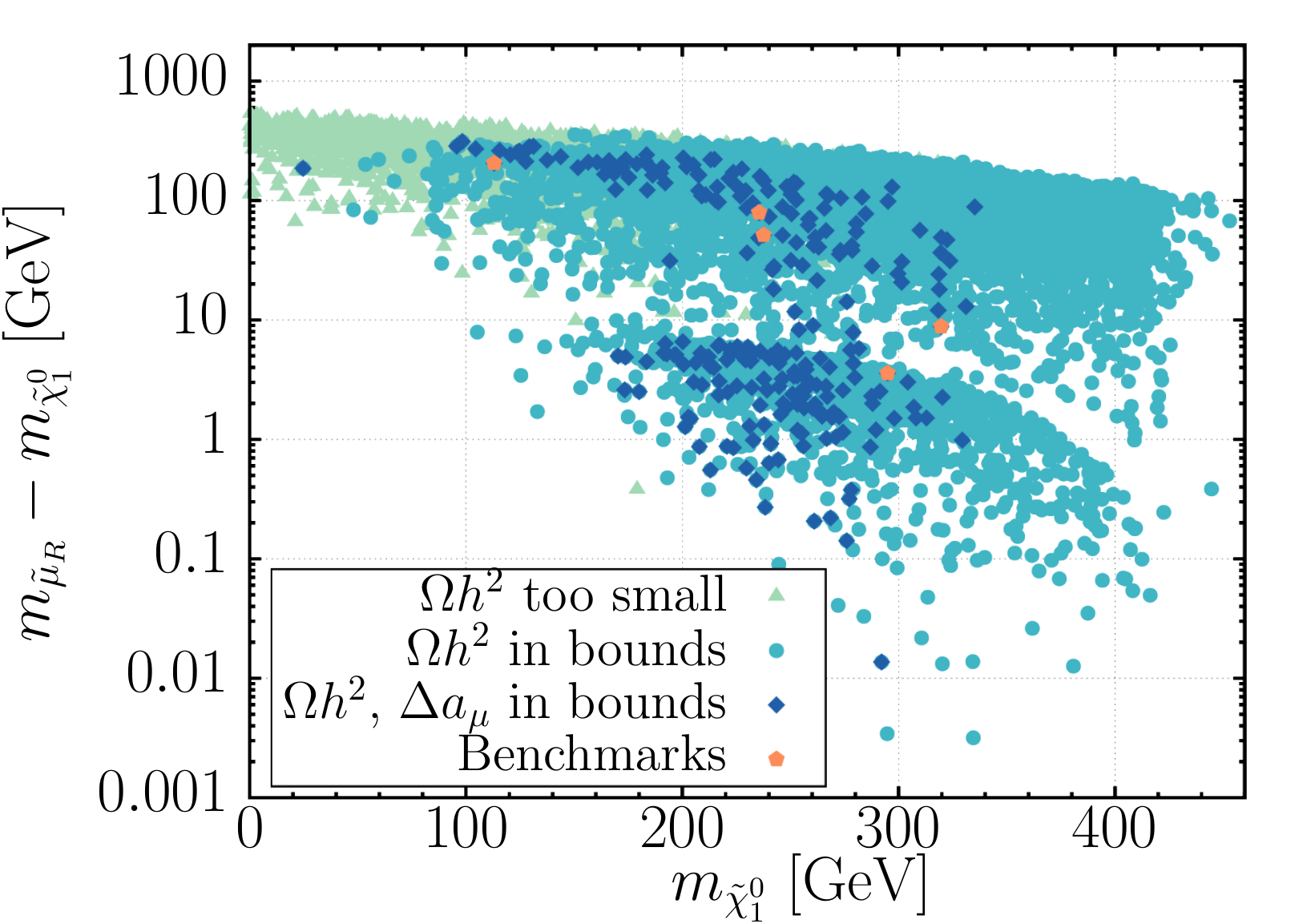}
  \caption{Mass gaps between the smuon and lightest neutralino masses $m_{\tilde{\mu}_{L/R}}$ and $m_{\tilde{\chi}^0_1}$. All dark blue diamonds are bino-like, whereas the light green triangles and turquoise circles are wino-like. The orange pentagons represent the benchmark points defined in table \ref{tab:benchmark_tab1}.}
  \label{fig:mass_gaps}
\end{figure}

\begin{figure}[H] 
  \centering
  \includegraphics[scale=.45]{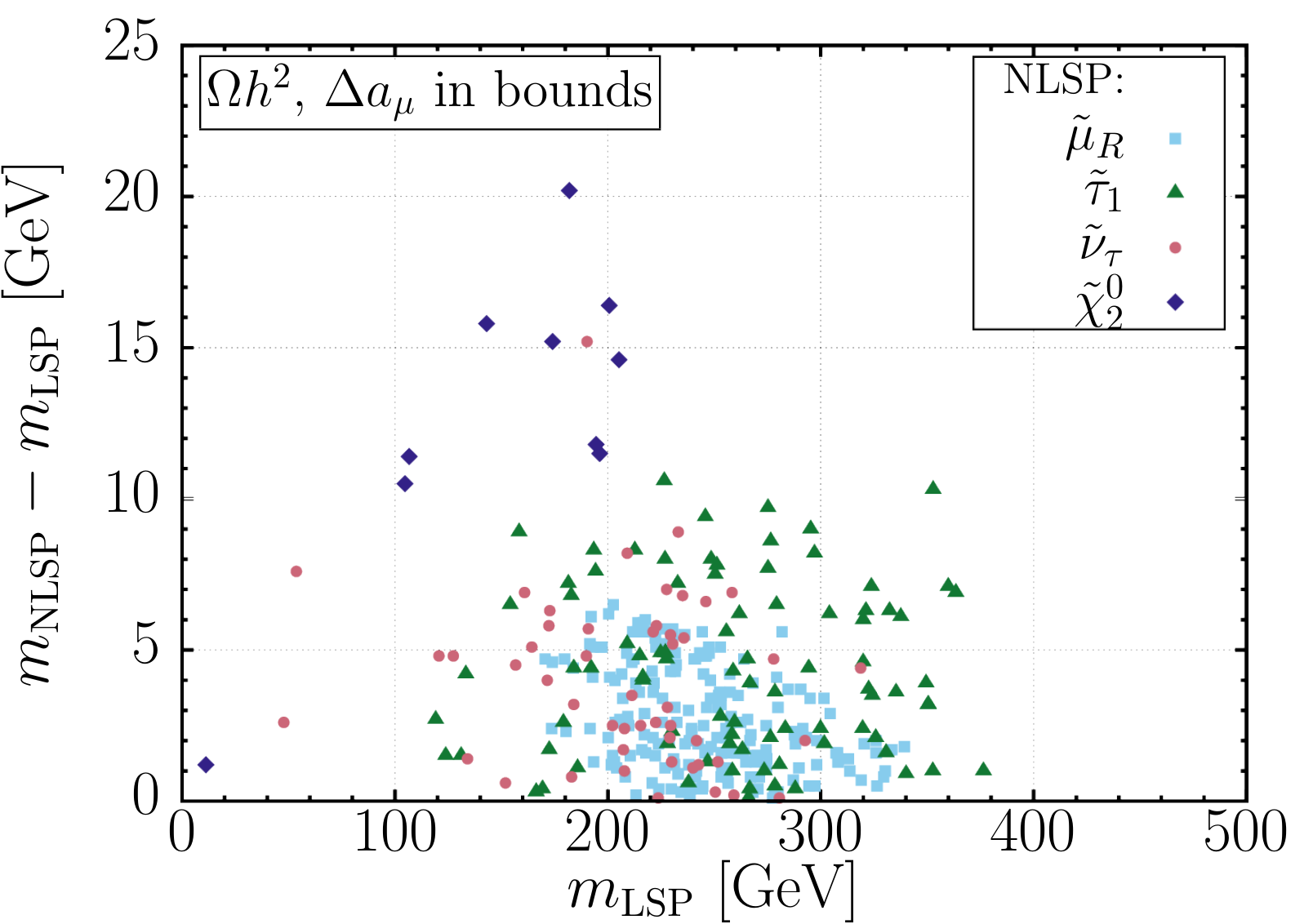}
\caption{Mass differences between the LSP and NLSP compared to the LSP mass. The top left plot has too small relic density, whereas the top right plot has the relic density in bounds and the lower central plot additionally has $\Delta a_\mu$ in bounds. For this plot, the LSP always is the lightest neutralino $\tilde{\chi}^0_1$.}
\label{fig:all_NLSP}
\end{figure}

\begin{figure}[H] 
  \centering
  \includegraphics[scale=.45]{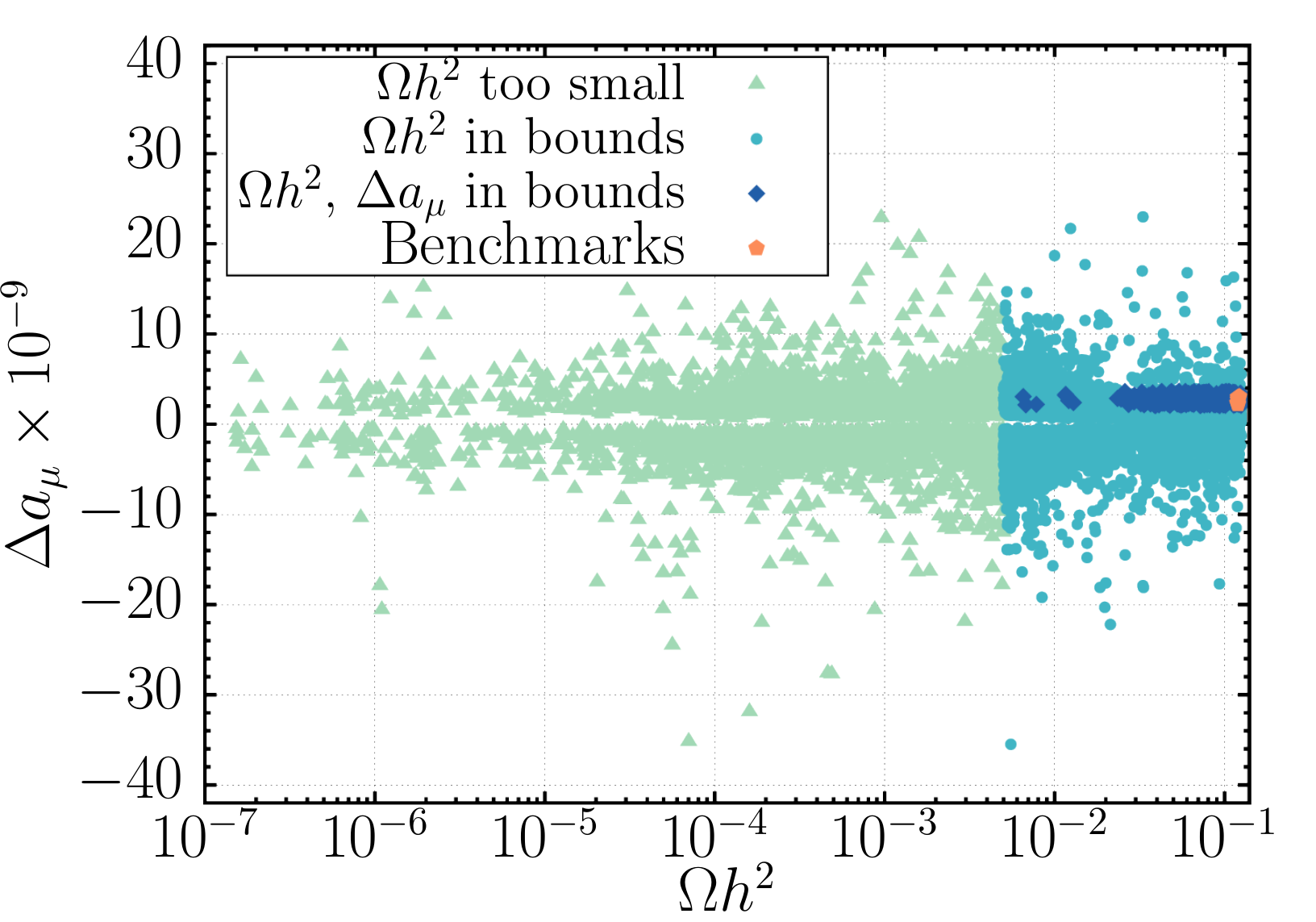}
  \includegraphics[scale=.45]{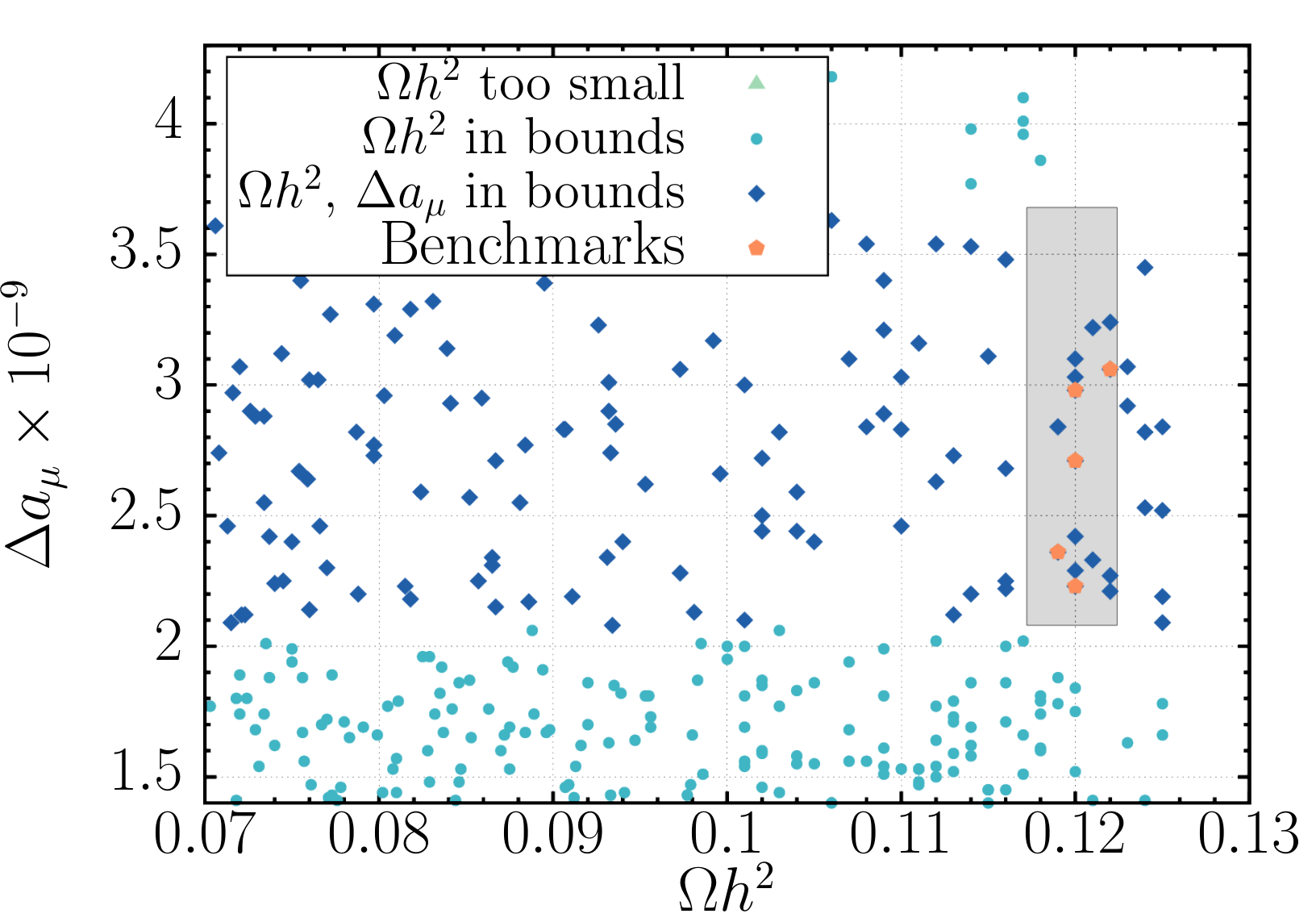}
  \caption{$\Delta a_\mu$ vs. $\Omega h^2$. The top left plot shows the full parameter spectrum, the top right plot a smaller excerpt of it with the grey shaded area being the 1$\sigma$ bound of $\Delta a_\mu$ and $\Omega h^2$. All dark blue diamonds are bino-like, whereas the light green triangles and turquoise circles are wino-like. The orange pentagons represent the benchmark points defined in table \ref{tab:benchmark_tab1}.}
  \label{fig:gmuon_vs_RD}
\end{figure}

It is clear that, in this case, diagram $(A)$ yields the main contribution to $\Delta a_\mu$. This is mainly due to the prefactor $\(\frac{M_1 \mu}{m^2_{\tilde{\mu}_L} m^2_{\tilde{\mu}_R}}\)$ from equation \ref{eq:loops:A} being large for large $\mu$ and small smuon masses. Additionally, eqauations \ref{eq:loops:B} \-- \ref{eq:loops:E} all feature $\mu$ in the denominator, thus leading to highly suppressed contributions from these diagrams.
In this scenario, dark matter is entirely bino-dominated for points with $\Delta a_\mu$ in the 1$\sigma$ bound (dark blue diamonds), leading to a viable relic density. This is visualised in figure \ref{fig:M1_vs_M2}.
In figure \ref{fig:mass_gaps}, we show the mass gaps between smuons and the LSP, which always is the lightest neutralino in this scenario. In case of left handed smuons, the mass gap for points featuring good $\Delta a_\mu$ and relic density (dark blue diamonds) is in a range of roughly 50 - 700 GeV, which is important for any collider phenomenology (cf. section \ref{sec:constraints}). As an example, muons emitted in the decay $\tilde{\mu}_L \to \tilde{\chi}^0_1 \, \mu_L$ would be very energetic, but most likely soft in the case of right handed smuons, as there are plenty of points with a mass gap below 50 GeV.
Figure \ref{fig:all_NLSP} shows the mass gaps between the LSP and NLSP vs. the LSP mass for too low relic density (top left plot), relic density in bounds (top right) and relic density as well as $\Delta a_\mu$ in bounds (bottom). In case of too low relic density, the first chargino is the NLSP for the majority of points and is degenerated in mass with the LSP. If the relic density increases, there are almost no chargino-NLSP's left and the NLSP changes to the right-handed smuon, but the first stauon and the $\tau$-sneutrino also yield significant amounts of NLSP's for this scenario. Also, all three of them are mass degenerated up to roughly 10 GeV with the LSP. In case of both relic density and $\Delta a_\mu$ being in the 1$\sigma$ bound, this picture does not change, but the favoured LSP mass is narrowed down from 100 \-- 400 GeV to 200 \-- 300 GeV for right-handed smuons. In case of $\tilde{\tau}_1$ or $\tilde{\nu}_\tau$, the LSP mass range is only slightly reduced.
\begin{figure}[H] 
  \centering
  \includegraphics[scale=.45]{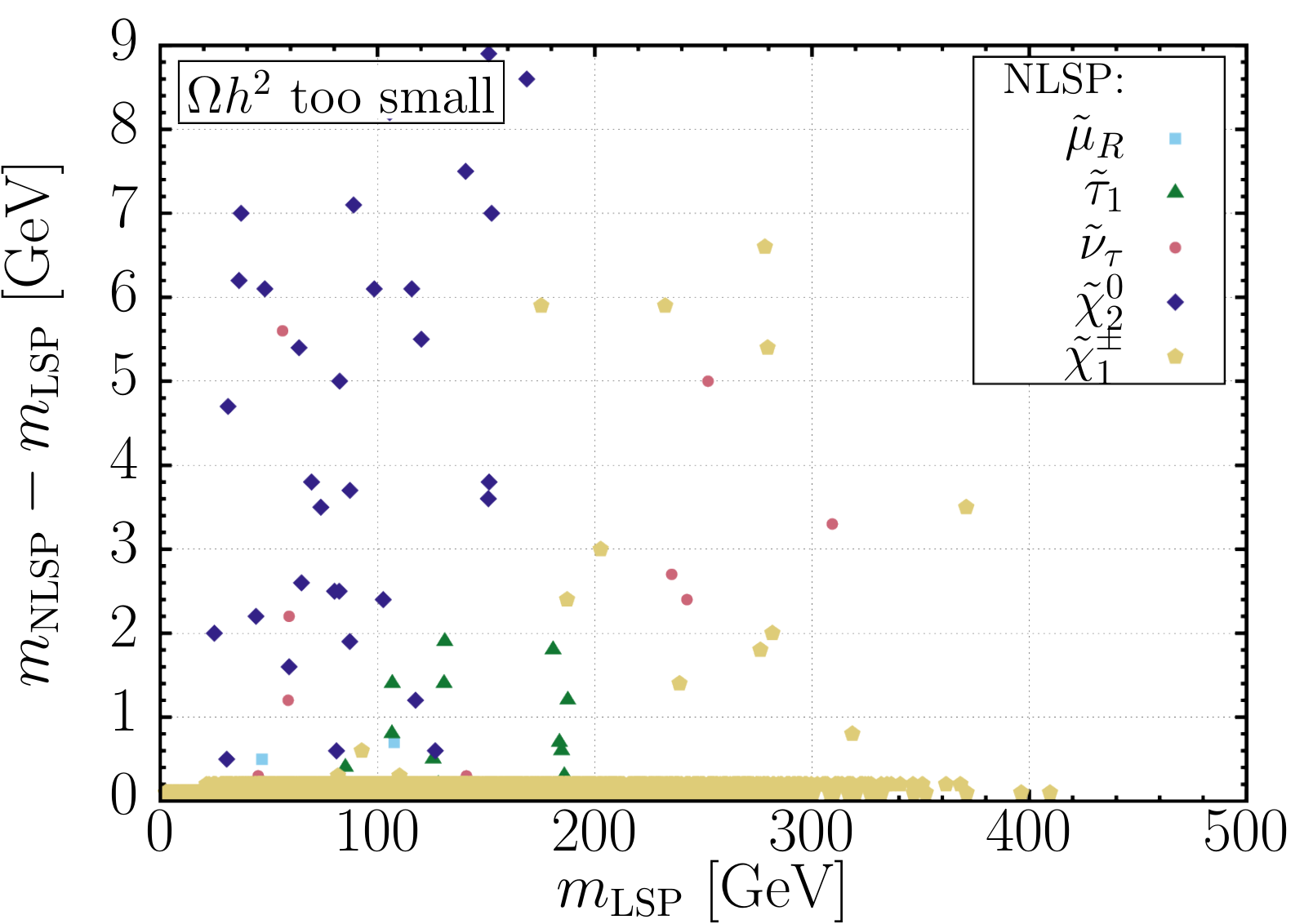}
  \includegraphics[scale=.45]{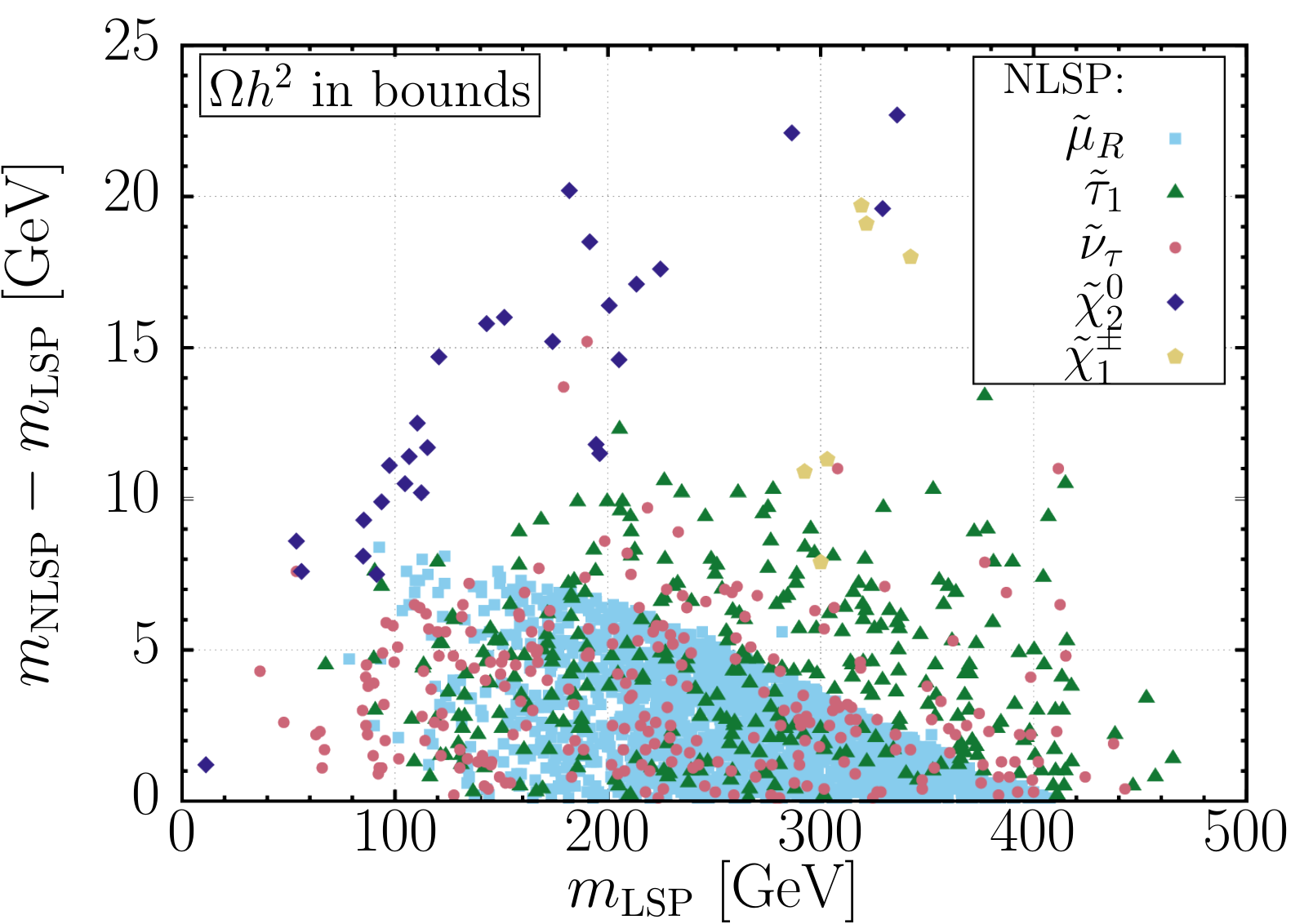}
\end{figure}
In figure \ref{fig:gmuon_vs_RD}, we show the $\Delta a_\mu \-- \Omega h^2$ plane and the respective 1$\sigma$ bounds as a grey shaded area. There are plenty of points lying close to the 1$\sigma$ bound w.r.t. $\Omega h^2$ and still many points in both 1$\sigma$ bounds. Based on the best points in the 1$\sigma$ bound (lower plot), we set up benchmark points (shown as orange pentagons in figures \ref{fig:diag_vs_mu} \-- \ref{fig:gmuon_vs_RD}) for the upcoming analysis for vacuum stability.
All benchmark points and their respective input parameters as well as a selection of the output parameters are shown below in table \ref{tab:benchmark_tab1}.
\begin{table}[H] 
	\centering
    \resizebox{\textwidth}{!}{%
	\begin{tabular}{cccccccccc}
\toprule
 & Benchmark: & BP6 & BP7 & BP8 & BP9 & BP10 &  \\ 
\midrule
\multirow{12}{*}{\rotatebox{90}{\textsc{Input at GUT scale}}} & $\tan \beta$ & 16.96 & 26.88 & 32.15 & 22.21 & 40.22 \\
 & sgn$(\mu)$       & +       & +       & +       & +       & +       & \\\cline{2-8}
 & $m_0$            & 238.8   & 149.6   & 106.5   & 271.5   & 137.5   & \multirow{10}{*}{\rotatebox{-90}{[GeV]}} \\
 & $m_1$            & 1426.7  & 1131.1  & 626.5   & 508.9   & 1470.7  & \\
 & $m_2$            & 239.2   & 302.7   & 125.3   & 193.5   & 178.4   & \\
 & $m_3$            & 1458.7  & 1631.9  & 1076.3  & 1434.2  & 1847.8  & \\
 & $M_1$            & 577.9   & 292.3   & 711.6   & 579.8   & 760.7   & \\
 & $M_2$            & 412.8   & 612.4   & 948.8   & -436.4  & 982.8   & \\
 & $M_3$            & 2195.7  & 2055.2  & 2680.5  & 2456.0  & 2524.6  & \\
 & $M_{h_1}$      & 670.6   & 2924.4  & 577.0   & 1512.8  & 1577.3  & \\
 & $M_{h_2}$      & 814.9   & 925.9   & 918.8   & 1306.2  & 1362.7  & \\
 & $A_{\text{tri}}$ & -2244.8 & -2776.6 & -1113.2 & -2896.2 & -2370.1 & \\
\hline
\multirow{24}{*}{\rotatebox{90}{\textsc{Masses}}} & $m_{h_0}$ & 124.1 & 124.1 & 123.5 & 124.5 & 123.6 & \multirow{26}{*}{\rotatebox{-90}{[GeV]}} \\
 & $m_{\tilde{g}}$          & 4595.1  & 4308.9  & 5497.4  & 5089.9  & 5201.7  &  \\
 & $m_{\tilde{q}^{1,2}_L}$  & 3931.0  & 3697.9  & 4709.6  & 4356.0  & 4468.8  &  \\
 & $m_{\tilde{b}_1}$        & 3527.8  & 3216.6  & 4257.5  & 3893.5  & 3878.2  &  \\
 & $m_{\tilde{t}_1}$        & 3412.9  & 3154.1  & 4068.0  & 3743.2  & 3842.3  &  \\
 & $m_{\tilde{q}^1_R}$      & 4183.9  & 3859.4  & 4731.4  & 4378.0  & 4683.5  &  \\
 & $m_{\tilde{q}^2_R}$      & 3936.0  & 3699.4  & 4690.4  & 4352.2  & 4445.5  &  \\
 & $m_{\tilde{b}_2}$        & 4137.3  & 3891.1  & 4637.2  & 4478.3  & 4510.4  &  \\
 & $m_{\tilde{t}_2}$        & 3586.5  & 3334.7  & 4286.0  & 3936.8  & 4038.8  &  \\
 & $m_{\tilde{e}_L}$        & 328.2   & 393.0   & 588.3   & 375.9   & 627.6   &  \\
 & $m_{\tilde{e}_R}$        & 1442.2  & 1136.2  & 684.3   & 552.4   & 1497.9  &  \\
 & $m_{\tilde{\mu}_L}$      & 328.2   & 393.0   & 588.1   & 375.9   & 627.7   &  \\
 & $m_{\tilde{\mu}_R}$      & 315.0   & 318.7   & 298.4   & 289.1   & 328.5   &  \\
 & $m_{\tilde{\tau}_1}$     & 248.1   & 120.0   & 485.0   & 244.9   & 328.2   &  \\
 & $m_{\tilde{\tau}_2}$     & 1445.0  & 1553.8  & 1052.4  & 1399.6  & 1720.5  &  \\
 & $m_{\tilde{\chi}^0_1}$   & 235.5   & 113.0   & 294.8   & 237.6   & 319.7   &  \\
 & $m_{\tilde{\chi}^0_2}$   & 310.4   & 483.2   & 758.4   & -426.1  & 792.1   &  \\
 & $m_{\tilde{\chi}^0_3}$   & -2942.2 & -2921.4 & -3116.3 & 3226.7  & -3273.1 &  \\
 & $m_{\tilde{\chi}^0_4}$   & 2942.6  & 2921.6  & 3116.9  & -3226.9 & 3273.5  &  \\
 & $m_{\tilde{\chi}^\pm_1}$ & 310.6   & 483.4   & 758.5   & 426.3   & 792.2   &  \\
 & $m_{\tilde{\chi}^\pm_2}$ & 2943.5  & 2922.6  & 3117.6  & 3227.8  & 3274.3  &  \\
 & $m_{\tilde{\nu}^e_L}$    & 318.5   & 384.8   & 582.7   & 367.4   & 622.4   &  \\
 & $m_{\tilde{\nu}^\mu_L}$  & 318.5   & 384.8   & 582.7   & 367.4   & 622.5   &  \\
 & $m_{\tilde{\nu}^\tau_L}$ & 243.3   & 129.8   & 517.2   & 247.0   & 350.5   &  \\\cline{1-7}
 & $Q$                      & 3409.7  & 3163.2  & 4072.1  & 3742.4  & 3845.1  &  \\
 & $\mu(Q)$                 & 2932.7  & 2917.6  & 3105.9  & 3217.7  & 3271.1  &  \\
\hline
\multirow{5}{*}{\rotatebox{90}{\textsc{Constraints}}} & Br$(b \to s \gamma)$      & $3.32 \times 10^{-4}$  & $3.29 \times 10^{-4}$  & $3.30 \times 10^{-4}$  & $3.32 \times 10^{-4}$  & $3.28 \times 10^{-4}$  &  \\
 & Br$(B_s \to \mu^+ \mu^-)$ & $3.07 \times 10^{-9}$  & $3.13 \times 10^{-9}$  & $3.14 \times 10^{-9}$  & $3.08 \times 10^{-9}$  & $3.32 \times 10^{-9}$  &  \\
 & $\sigma^{\text{DD SI}}$   & $9.69 \times 10^{-13}$ & $4.44 \times 10^{-13}$ & $6.65 \times 10^{-13}$ & $5.50 \times 10^{-13}$ & $6.31 \times 10^{-13}$ & [pb]  \\
 & $\Omega h^2$              & $1.20 \times 10^{-1}$  & $1.22 \times 10^{-1}$  & $1.20 \times 10^{-1}$  & $1.20 \times 10^{-1}$  & $1.19 \times 10^{-1}$  &  \\
 & $\Delta a_\mu$            & $2.71 \times 10^{-9}$  & $3.06 \times 10^{-9}$  & $2.23 \times 10^{-9}$  & $2.98 \times 10^{-9}$  & $2.36 \times 10^{-9}$  &  \\
\bottomrule
    \end{tabular}}
    \caption{Input and Output parameters for the benchmark points with the most accurate $\Delta a_\mu$ and $\Omega h^2$ in the case of large $\mu(Q)$ and all other constraints being fulfilled. $\tilde{q}^{i}$ labels the $i$-th generation of squarks.}
    \label{tab:benchmark_tab1}
\end{table}

%% file: 06-vacuum-stability.tex
\section{Vacuum Stability}
\label{sec:vacuum-stability}

\texttt{SoftSUSY} implements two-loop tadpole contributions to the minimization conditions to ensure the breaking of electroweak symmetry by Higgs VEVs. As with other spectrum generators, the minimization conditions are used to fix parameters of the theory in such a way that the desired vacuum is a minimum of the scalar potential. One downside of this procedure is that other solutions to the minimization conditions might exist and lie lower in the scalar potential of the theory. At the same time, color- and charge- breaking (CCB) VEVs are usually ignored and such minima might also exist and lie lower than the desired vacuum. 

It is then interesting to understand if the points in our scans suffer from CCB minima, whether they are lower than the desired vacuum and in that case if the desired vacuum is sufficiently long-lived (meta-stable). Although approximate analytical conditions for the avoidance of CCB minima exist for the MSSM, a full numerical study of the one-loop effective potential is often needed as the conditions are neither sufficient nor necessary to ensure the absence of such minima \cite{Camargo-Molina:2013sta}. In addition, such analytical rules are based on a tree-level analysis and are thus irrelevant for points where the symmetry breaking occurs only at one-loop. 

Using \texttt{Vevacious} \cite{Camargo-Molina:2013qva} we performed a numerical analysis of the tree and one-loop effective potential for a set of benchmark parameter points allowing for stop and stau VEVs. Due to the fact that the desired vacuum comes as a solution of two-loop minimization conditions we found that quite often the EWSB minimum only appears after two-loop contributions to the effective potential are considered. For such parameter points an analysis with \texttt{Vevacious}, which uses the one-loop effective potential, was not possible and thus the vacuum stability analysis was inconclusive. However, it was still possible to find parameter points where the EWSB minimum (the desired vacuum) develops at tree-level or one-loop. In the case of minima appearing only at one-loop, a careful numerical minimization of the one-loop effective potential was required, as \texttt{Vevacious} uses the tree-level minima as starting points for numerical minimization therefore missing such cases out of the box. It was possible however to study the vacuum stability in a point by point basis by starting the numerical minimization around the field values for the EWSB minimum that develops once two-loop contributions are considered. 

\begin{figure}[H] 
  \centering
  \includegraphics[width=0.49\linewidth]{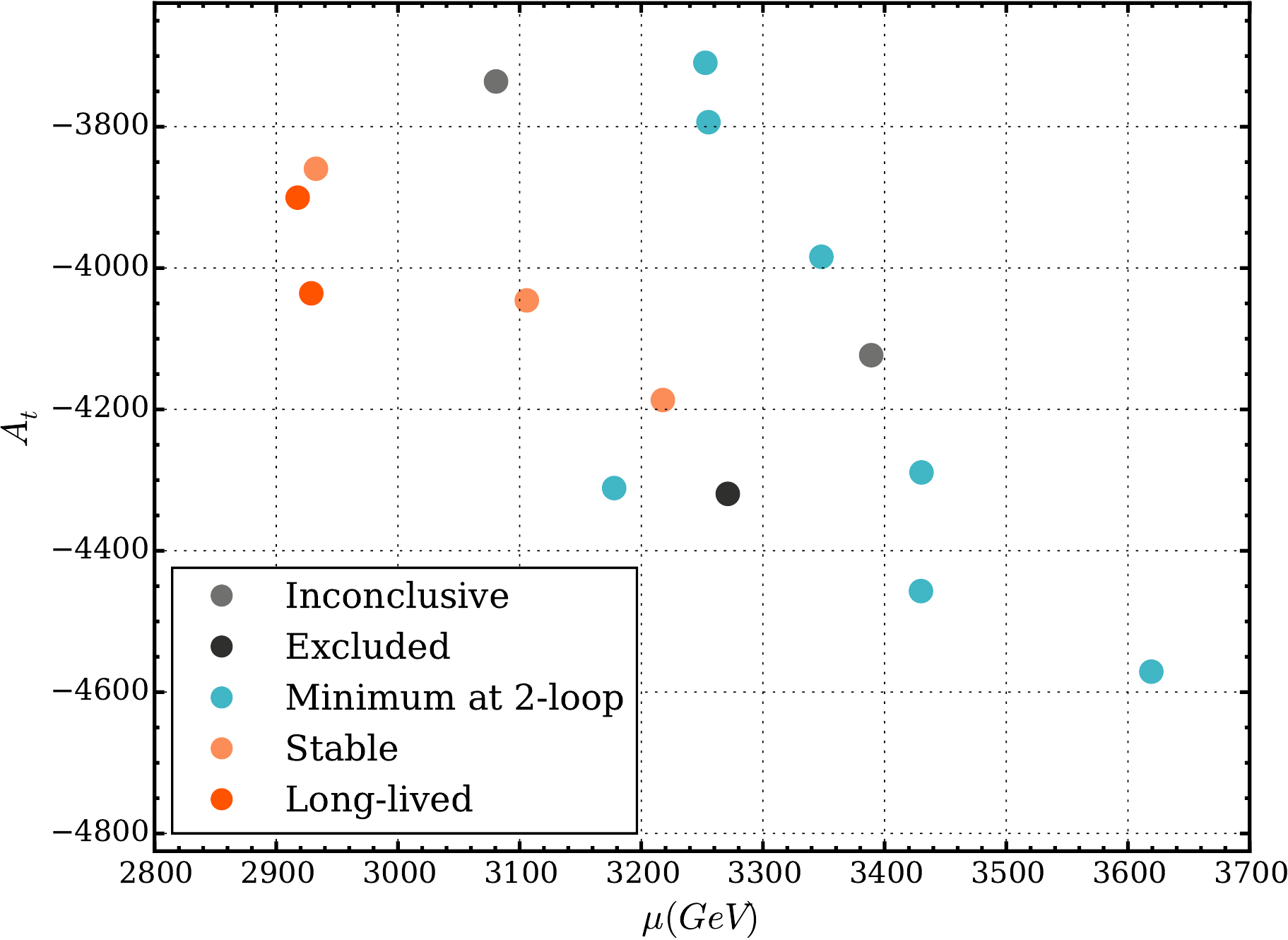}
    \includegraphics[width=0.463\linewidth]{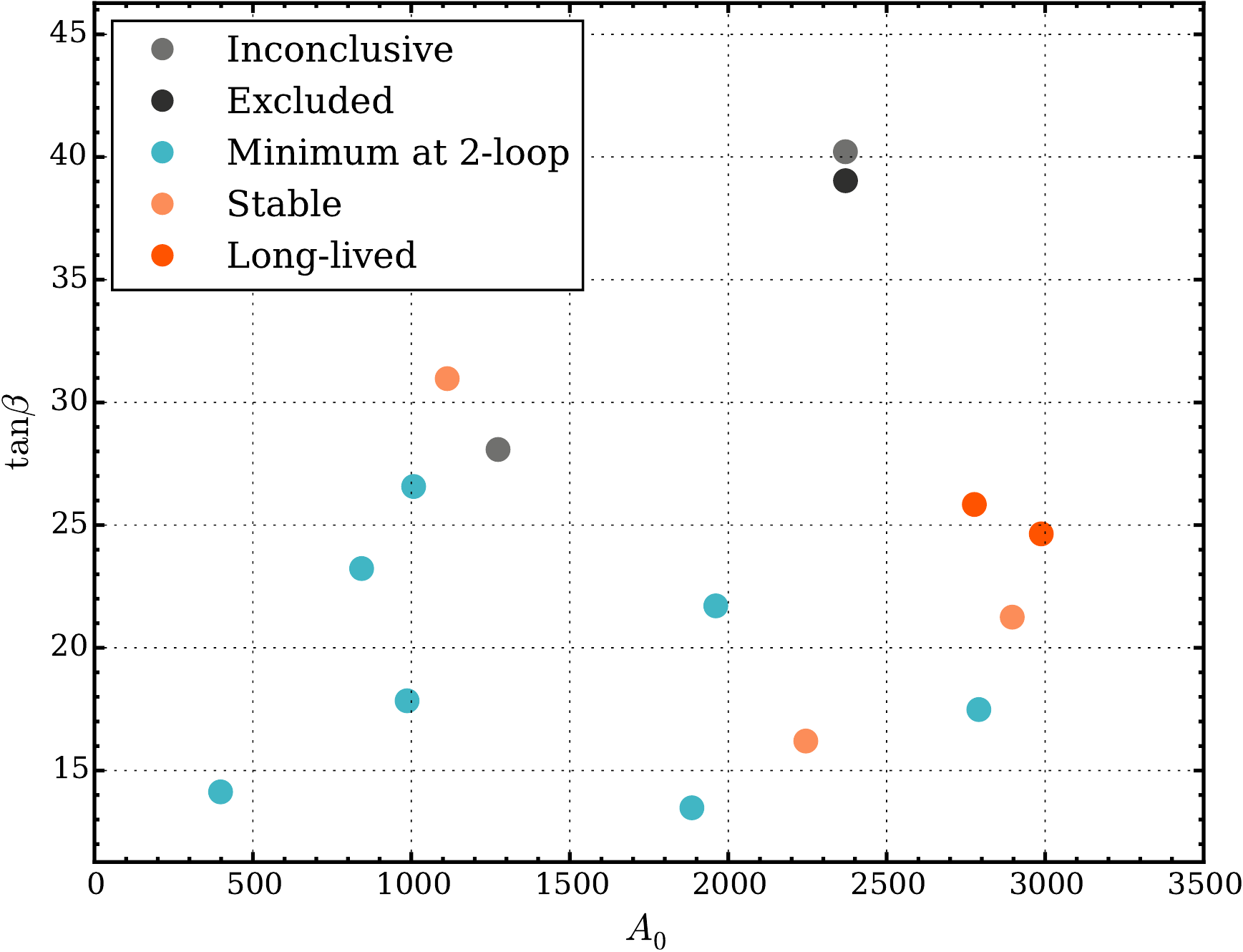}
  \caption{Vacuum stability analysis for set of points fulfilling all other constraints in the large-$\mu$ region: Orange and red points correspond to the final benchmark points for which the desired vacua are stable or long-lived respectively. For light blue points (labeled ``Minimum at two-loop'' ) the desired vacuum appears only when two-loop corrections are included and thus could not be studied with current tools. Black points showed deeper CCB minima with $< 1 \%$ survival probability of the desired vacuum. Gray points showed CCB minima at 1-loop but the desired vacua appear after two-loop corrections are included. }
  \label{fig:VS_mu_vs_tanb}
\end{figure}

For the points considered in section \ref{subsec:small_mu} and shown in table \ref{tab:benchmark_tab_small}, the desired vacuum was the global minimum of the one-loop effective potential. For the points considered in section \ref{subsec:large_mu}, we started with a set of benchmark points satisfying all the constraints considered in the previous sections, and after performing the vacuum stability analysis we selected those where the desired vacuum was either the global minimum (and thus stable) or long-lived after considering tunneling to deeper CCB minima at zero and non-zero temperature. The result of the analysis is shown in figure \ref{fig:VS_mu_vs_tanb}. In this figure we can see that points with larger $\mu$ roughly correspond to those for which the desired vacuum develops once two-loop corrections are considered, as could be naively expected. In addition the stable and long-lived points tend to have larger $|A_t|$ and $A_0$ together with lower $\tan \beta$ values. This comes from the fact that the larger $\tan \beta$ is the smaller $m_{\tilde{\tau}_R}^2 $, increasing the chance for $\tilde{\tau}$ VEVs. Conversely, lower $\tan \beta$ thus allows for higher values of $A_0$ and $|A_t|$ a combination that allows the points to fulfill all other constrains together with vacuum stability. The points for which the desired vacuum was either global or long-lived minimum (red and orange in  fig. \ref{fig:VS_mu_vs_tanb}) correspond to the benchmark points shown in light orange in figures \ref{fig:M1_vs_M2}, \ref{fig:mass_gaps} and \ref{fig:gmuon_vs_RD}. 


%

%% file: 07-conclusions.tex
\section{Conclusions} \label{sec:conclusion}

The anomalous magnetic moment of the muon continues to show a disagreement with the SM which suggests new physics at a relatively low mass scale. The leading candidate for such new physics is the MSSM with light sleptons and light charginos and neutralinos, which can contribute substantially to $\Delta a_\mu$ at one-loop and explain the experimental $\Delta a_\mu$ measurements. Such a SUSY spectrum as low as a few hundreds GeV requires to explain $\Delta a_\mu$ contrasts with the failure of the LHC to discover coloured superpartners such as squarks and gluinos, leading to stringent bounds on such sparticles, requiring their masses to typically lie above the TeV scale. The Higgs boson mass also requires at least some stop masses above the TeV scale in the MSSM. 

From the experimental side, these constraints are not inconsistent with having light sleptons and gauginos down to about 100 GeV, since the LHC sensitivity to colour singlets is significantly lower than to coloured particles. At the same time, from the theory side it is very hard to accommodate light sleptons and heavy squarks for all generations at the weak scale in the cMSSM or mSUGRA model with universal sfermion masses at the GUT scale. This is especially difficult if one takes into account combined collider and non-collider constraints including those from the dark matter relic density.

Such a tension strongly favours MSSM models with non-universal sfermion masses at the GUT scale like the pMSSM which relax the constraints of the CMSSM without introducing excessive flavour changing neutral currents and without unleashing all the 100 or so parameters of the MSSM. However, the pMSSM still contains 19 SUSY parameters and is not particularly well theoretically motivated. In this paper, we have considered a theoretically very well motivated scenario, which involves just four soft scalar masses at the GUT scale, namely $m_0$ (a universal left-handed scalar mass) and $m_1$, $m_2$, $m_3$ (three universal right-handed scalar masses, one for each family), together with non-universal gaugino and trilinear soft masses. In this model, the first and second family sleptons can be light to explain $\Delta a_\mu$ while simultaneously, $m_3$ can be large enough to provide enough mass for the Higgs boson and the agreement with other observables such as $\text{Br}(b \to s \gamma)$ and $\text{Br}(B_s \to \mu^+ \mu^-)$ stays valid.

The comprehensive scan over the soft parameter space of the model, exploiting the relatively small number of soft input masses (as compared for example to the pMSSM), has confirmed the existence of viable points which satisfy both $\Delta a_\mu$ and dark matter constaints neatly dividing into two sets: small $\mu$ and large $\mu$, which we subsequently investigated in detail separately. For these two parameter regions, we were able to understand the dominant effects leading to successful $\Delta a_\mu$ as well as the characteristics of the dark matter candidate, while satisfying all other experimental constraints. For example we investigated the NLSP to understand which SUSY particle is responsible for the effective co-annihilation as well as the LSP-NLSP mass splitting, which is very important experimentally. We also proposed sets of benchmark points for each scenario and checked the vacuum stability for all benchmark points, especially for the large $\mu$ case where vacuum stability is an issue.

The small $\mu \lesssim 400$ GeV region involves a bino-like neutralino LSP which annihilates in the early Universe either resonantly, if its
mass is around half the mass of the $Z$ or Higgs boson, or via co-annihilation with the higgsino states if the $\mu$ parameter is about 15 GeV higher than the LSP mass. The benchmarks are chosen such that there is a large mass gap of around 100 GeV between the LSP and the smuon mass, so that the smuon decay will involve a hard muon, providing a clear signal at the LHC. For all these small $\mu$ cases, $\Delta a_\mu$ is dominated by diagrams (B) and (E) of figure~\ref{fig:1-loop-amu}. The large $\mu \sim 3$ TeV region also involves a bino-like neutralino LSP which co-annihilates in the early Universe with an NLSP which may be $\tilde{\tau}_1$, $\tilde{\nu}_{\tau}$, $\tilde{\chi}_2$ or $\tilde{\mu}$, depending on the precise parameters. In all these cases, the dominant contribution to $\Delta a_\mu$ comes from diagram (A) of figure~\ref{fig:1-loop-amu}. In both scenarios, heavy gluinos (above 2 TeV) help to split the squark and slepton masses of the first two generations, yielding heavy squark masses satisfying the LHC bounds on the first and second family squarks, while allowing light sleptons. These scenarios both predict light smuons (100-300 GeV), which can be probed via leptonic signatures and even potentially explain di-lepton excesses reported by the ATLAS and CMS collaborations. In addition, the  small $\mu$ scenario also predicts quite light charginos and second neutralinos exhibiting 
di-lepton or tri-lepton signatures which can be tested in the near future and/or explain the di-lepton excesses mentioned above.

In conclusion, the MSSM with a Pati-Salam gauge group broken at the GUT scale and flavour symmetries $A_4$ and $Z_5$, which unify the soft masses of the left-handed (but not right-handed) squarks and sleptons, provides a well motivated framework with a relatively low number of input soft masses, which is capable of accounting for $\Delta a_\mu$ as well as providing good dark matter candidates, consistently with all other experimental and theoretical constraints. We emphasise that (unlike some other models) the A to Z Pati-Salam model initially was not designed to explain $\Delta a_\mu$, since its primary motivation was to explain the flavour mass and mixing of quarks and leptons, in particular neutrinos. Nevertheless, we have seen that the model is well suited to account for $\Delta a_\mu$, while simultaneously providing a good dark matter candidate, namely the lightest neutralino, which is consistently bino-like in nature. The characteristic SUSY spectra presented here should enable this model to be distinguished from other less well motivated models such as the pMSSM.